\documentclass[aps,pra,twocolumn,showpacs,floatfix]{revtex4}
 \usepackage{epsfig,amsmath,amssymb,bm,epsf,graphics}

\def\rlo{{\lambda}}
\def\rtr{{\tau}}

\def\ers{{\epsilon}}

\def\dsl{{\cal N}}
\def\cw{{\cal W}}
\def\mw{{w}}
\def\op{{\Omega}}
\def\den{c}
\def\hrs{{\rm HRS}}
\def\ors{{\rm 1RS}}
\def\zrs{{\rm 0RS}}
\def\newmu{m}
\def\stiff{{\mu}}
\def\rpf{{\cal Z}}
\def\fim{{\cal D}}
\def\DPS{\displaystyle}
\def\lsz{\xi_{0}}
\def\dbar{{\mathchar'26\mkern-12mud}}   

\def\GatZ{\Gamma}
\def\wh{\widehat}
\def\nh{\hat}
\def\wt{\widetilde}
\def\cpd{{\cal P}}
\def\bfgamma{{\bf g}}
\def\lpc{{\ell_{\rm p}}}
\def\lbar{{\mathchar'26\mkern-10mu\ell}}   
\def\coshort{\ell_{<}}
\def\colong{\ell_{>}}
\def\cbshort{\lbar_{<}}
\def\cblong{\lbar_{>}}
\def\cRshort{{\widetilde{\ell}}_{<}}
\def\linsize{{\cal L}}
\def\smhalf{{\scriptstyle\frac{\scriptstyle 1}{\scriptstyle 2}}}
\def\smquar{{\scriptstyle\frac{\scriptstyle 1}{\scriptstyle 4}}}
  \def\Jinver{\frac{1}{J}}

\def\sdcon{{\bf v}}
\def\llshift{{\xi}_{\rm fl}}
\def\lltyp{{\xi}_{\rm typ}}
\def\cloud{{\wp}}
\def\mypercent{{\%}}
\def\GMZeta{{\eta}}
\def\qlf{\widetilde{Q}}
\begin{document}
\title{Goldstone-type fluctuations and their 
implications for the amorphous solid state}

\author{%
Paul M. Goldbart\rlap,$^{(a,b)}$
Swagatam Mukhopadhyay$^{(a)}$ and
Annette Zippelius$^{(b,c)}$}
\affiliation{%
$^{(a)}$Department of Physics,
University of Illinois at Urbana-Champaign,
1110 West Green Street, Urbana, Illinois 61801-3080, U.S.A.
\\
$^{(b)}$Kavli Institute for Theoretical Physics,
University of California--Santa Barbara, California 93106, U.S.A.
\\
$^{(c)}$Institut f\"ur Theoretische Physik,
Universit\"at G\"ottingen, D-37073 G\"ottingen, Germany%
}

 \date{March 21, 2004}

\begin{abstract}
In sufficiently high spatial dimensions, 
the formation of the amorphous (i.e.~random) solid state of matter, 
e.g., upon sufficent crosslinking of a macromolecular fluid, involves 
particle localization and, concommitantly, the spontaneous 
breakdown of the (global, continuous) symmetry of translations.  
Correspondingly, the state supports Goldstone-type low energy, 
long wave-length fluctuations, the structure and implications 
of which are identified and explored from the perspective of 
an appropriate replica field theory.  
In terms of this replica perspective, the lost symmetry is that 
of relative translations of the replicas; common translations 
remain as intact symmetries, reflecting the statistical homogeneity 
of the amorphous solid state.
What emerges is a picture of the Goldstone-type fluctuations of 
the amorphous solid state as shear deformations of an elastic 
medium, along with a derivation of the shear modulus and the 
elastic free energy of the state.  
The consequences of these fluctuations---which dominate deep 
inside the amorphous solid state---for the order parameter 
of the amorphous solid state are ascertained and interpreted 
in terms of their impact on the statistical distribution of 
localization lengths, a central diagnostic of the the state. 
The correlations of these order parameter fluctuations are also 
determined, and are shown to contain information concerning 
further diagnostics of the amorphous solid state, such as spatial 
correlations in the statistics of the localization characteristics. 
Special attention is paid to the properties of the amorphous solid state 
in two spatial dimensions, for which it is shown that Goldstone-type 
fluctuations destroy particle localization, the order parameter is 
driven to zero, and power-law order-parameter correlations hold. 
%
\end{abstract}
\pacs{61.43.-j, 82.70.Gg, 64.60.Ak}
%
%

\maketitle

\section{Introduction}
\label{sec:intro}
The aim of this Paper is to identify the long wave-length, 
low energy fluctuations of the amorphous (i.e.~random) solid state, 
and to investigate their physical consequences.  In particular, by
constructing an effective free energy that governs these
{\it Goldstone-type\/} fluctuations, we shall determine the 
elastic properties of the amorphous solid, including its static shear
modulus.  We shall also analyze the effect of these fluctuations
on the amorphous solid order parameter, and hence determine their
impact on physical quantities such as the distribution of localization 
lengths and the order parameter correlations.  Along the way we shall 
reveal the physical information encoded in these correlations.  The 
treatment presented here is valid deep inside the amorphous solid 
state, i.e., far from the critical point associated with the phase 
transition from the liquid to this state.  Fluctuation phenomena
such as those discussed in the present Paper cannot be captured by 
a solely percolative approach to amorphous solidification.  
A brief account of the work reported in the present Paper has been 
given elsewhere~\cite{Mukhopadhyayetal2003}.

We shall pay particular attention to systems of spatial dimension
two, for which we shall see that the effect of fluctuations is
strong: particle localization is destroyed, the order parameter is
driven to zero, and order-parameter correlations decay as a power 
law in the separation between points in the sample.  Thus we shall 
see that the amorphous solid state is, in many respects, similar to 
other states of matter exhibiting (or nearly exhibiting) spontaneously 
broken continuous symmetry.  

This Paper is organized as follows. 
In Sec.~\ref{sec:RSOPSSB} we sketch the properties of the amorphous solid 
order parameter, including the form it takes in the amorphous solid state, 
focusing on symmetry properties and how they manifest themselves within 
the replica formalism. 
In Sec.~\ref{sec:GoldstoneStructure} we describe the structure of 
the low energy, long wave-length Goldstone-type excitations of the 
amorphous solid state in terms of distortions of the value of 
the order parameter.  Here, we also make the identification of these 
order-parameter distortions as local displacements of the 
amorphous solid. 
In Secs.~\ref{sec:GoldstoneEnergetics} and \ref{sec:ShearModulus}
we determine the energetics of these Goldstone-type excitations by  
beginning with a Landau-type free energy expressed in terms of the 
amorphous solid order parameter and ending with elasticity theory.  
Along the way, we derive a formula for the elastic shear modulus, 
which shows how this modulus vanishes, as the liquid state is 
approached, at the classical (i.e.~mean-field theory) level.
In Sec.~\ref{sec:OPreduction} we discuss the impact of Goldstone-type 
fluctuations on the structure of the amorphous solid state by examining 
how they diminish the order parameter and modify the distribution of 
localization lengths.  We  analyze the impact of such fluctuations on 
the order-parameter correlations, and also catalog the various 
length-scales that feature in the Paper.
In Sec.~\ref{sec:TwoDimensionRS} we take a closer look at the effects  
of Goldstone-type fluctuations on structure and correlations in 
two-dimensional amorphous solids.  In particular, we show that 
Goldstone-type fluctuations destroy particle localization, the order 
parameter is driven to zero, and power-law order-parameter correlations 
hold, and we illustrate these features for certain special cases.
In Sec.~\ref{sec:OPcorrelators} we discuss the physical content of 
order-parameter correlations in terms of spatial correlations in the 
statistics of the localization lengths.  We also introduce 
distributions of correlators, and relate their moments to 
order-parameter correlators. 
Some concluding remarks are given in Sec.~\ref{sec:conclusions}.  
Technical details are relegated to four appendices.

\section{Amorphous solid order parameter, 
         symmetries and symmetry breaking}
\label{sec:RSOPSSB}
\subsection{Order parameter}
\label{sec:RSO}
The solid state of randomly and permanently constrained matter,
well exemplified by the rubbery state of vulcanized macromolecular matter,
may be detected and characterized via the following order parameter, which
depends on an arbitrary number 
$\nu$ ($\geq 2$) of tunable (but nonzero)
wave vectors $\{{\bf k}^{1},{\bf k}^{2},\ldots,{\bf k}^{\nu}\}$:

\def\OPnonrep{{\widetilde{\Omega}}}
\begin{equation}
\OPnonrep
({\bf k}^{1},\ldots{\bf k}^{\nu})=
\Big[\frac{1}{J}
\sum_{j=1}^J
\langle{\rm e}^{i{\bf k}^{1}\cdot{\bf R}_{j}}\rangle
\langle{\rm e}^{i{\bf k}^{2}\cdot{\bf R}_{j}}\rangle
\cdots
\langle{\rm e}^{i{\bf k}^{\nu}\cdot{\bf R}_{j}}\rangle
\Big].
\label{eq:phyop}
\end{equation}
Here, the vectors $\{{\bf R}_{j}\}_{j=1}^{J}$ give the positions of the
$J$ particles that constitute the system; they inhabit
a large, $D$-dimensional, hyper-cubic region ${\cal V}$ of volume $V$.
Angular brackets $\langle\cdots\rangle$ indicate an equilibrium expectation
value taken in the presence of a given realization of the quenched
random constraints, possibly in a state with spontaneously broken symmetry.
Square brackets $[\cdots]$ denote an average over the number and
specifications of the quenched random constraints.  Periodic boundary
conditions discretize the wave vectors ${\bf k}$ to the values 
$2\pi{\bf m}/V^{1/D}$ in terms of $D$-tuples of integers ${\bf m}$.
Additional discussion of this circle of ideas is given in 
Refs.~\cite{GoldbartAdvPhy1996,GoldbartTrieste2000}. 
 
The order parameter~(\ref{eq:phyop}) detects and diagnoses the 
amorphous solid state by sensing and quantifying the presence 
of static random waves in the particle density, as implied by 
$\langle{\rm e}^{i{\bf k}\cdot{\bf R}_{j}}\rangle\neq 0$ 
(for ${\bf k}\ne{\bf 0}$).  
To acquire some feeling for how it works, consider the
illustrative example in which a fraction $1-Q$ of the particles 
are unlocalized whilst the remaining fraction $Q$ are localized,
harmonically and isotropically but randomly, having random mean
positions $\langle{\bf R}_{j}\rangle$ and random mean-square
displacements from those positions
\begin{equation}
\left\langle
\left(
{\bf R}_{j}-\langle{\bf R}_{j}\rangle
\right)_{d}\,
\left(
{\bf R}_{j}-\langle{\bf R}_{j}\rangle
\right)_{d^{\prime}}
\right\rangle
=\delta_{dd^{\prime}}\,\xi_{j}^{2}, 
\label{eq:RMSflucts}
\end{equation}
where $d$ and $d^{\prime}$ are cartesian indices running from 
$1$ to $D$.  It is straightforward to see that for this example 
the order parameter becomes
\begin{eqnarray}
&&
\OPnonrep
({\bf k}^{1},\ldots{\bf k}^{\nu})
\nonumber\\
&&= 
Q\,\delta_{{\bf 0},\sum_{a=1}^{\nu}{\bf k}^{a}}
\!\int_{0}^{\infty}  \!\!\!\!\! d\xi^{2}\dsl(\xi^{2})\,
\exp\Big(-\frac{\xi^{2}}{2}
\sum\nolimits_{a=1}^{\nu}\vert{\bf k}^{a}\vert^{2}\Big)\nonumber,
\label{eq:interpretop}
\end{eqnarray}
where
\begin{equation}
\dsl(\xi^{2})\equiv
\left[
(QJ)^{-1}
\sum\nolimits_{j\,{\rm loc.}}\delta(\xi^{2}-\xi_{j}^{2})
\right]
\label{eq:defdistrib}
\end{equation}
is the dis{\-}order-averaged distribution of squared 
localization lengths $\xi^{2}$ of the localized fraction of particles~\cite{Castillo+Goldbart+Zippelius1994,GoldbartAdvPhy1996,GoldbartTrieste2000}.

The illustrative example correctly captures the pattern in which 
symmetry is spontaneously broken when there are enough random constraints 
to produce the amorphous solid state: microscopically, random localization 
fully eliminates translational symmetry; but macroscopically this 
elimination is not evident.  Owing to the absence of any residual symmetry, 
such as the discrete translational symmetry of crystallinity, all
macroscopic observables are those of a translationally invariant system.  
This shows up as the vanishing of the order parameter, even in the amorphous 
solid state, unless the wave vectors sum to zero, i.e., 
$\sum_{a=1}^{\nu}{\bf k}^{a}={\bf 0}$.

The case $\nu=1$ is excluded from the list of order parameter 
components shown in Eq.~(\ref{eq:phyop}).  This case corresponds 
to {\it macroscopic\/} density fluctuations, and these are assumed 
to remain small and stable (i.e.~non-critical) near the amorphous 
solidification transition, being suppressed by forces, such as the 
excluded-volume interaction, that tend to maintain homogeneity.  
Additional insight into the nature of the constraint-induced instability 
of the liquid state and its resolution (in terms of the formation 
of the amorphous solid state---a mechanism for evading macroscopic 
density fluctuations) is given in Ref.~\cite{GoldbartTrieste2000}, 
especially Sec.~4.2.

\subsection{Symmetries and symmetry breaking; replica formulation}
\label{sec:symmetries}
How does the order parameter~(\ref{eq:phyop}) transform under
translations of the particles?  If, in the element $\langle\exp i{\bf
  k}^{a}\cdot{\bf R}_{j}\rangle$, one makes the translation ${\bf
  R}_{j}\to{\bf R}_{j}+{\bf r}^{a}$ then the element is multiplied by
a factor $\exp i{\bf k}^{a}\cdot{\bf r}^{a}$ and so the order
parameter acquires a factor $\exp i\sum_{a=1}^{\nu}{\bf
  k}^{a}\cdot{\bf r}^{a}$.  Now, in the fluid state no particles are
localized and the order parameter has the value zero, so it is
invariant under the aforementioned translations.  By contrast, in the
amorphous solid state the order parameter is nonzero, provided the wave
vectors sum to zero.  Thus, the order parameter varies under the
translations unless the translation is common to each element, i.e.,
${\bf r}^{a}={\bf r}$.
To summarize, the liquid-state symmetry of the independent translations of 
the elements is broken down, at the amorphous solidification transition, 
to the residual symmetry of the common translation of the elements.

When replicas are employed to perform the average over the quenched
disorder (i.e.~the number and location of the constraints), what
emerges is a theory of a field $\wh{\Omega}(x)$ defined over
$(1+n)$-fold replicated space $x$, so that the argument $x$ means 
the collection of $(1+n)$ position $D$-vectors 
$\{{\bf x}^{0},{\bf x}^{1},\ldots,{\bf x}^{n}\}$ 
conjugate to the wave vectors 
$\{{\bf k}^{0},{\bf k}^{1},\ldots,{\bf k}^{n}\}$.  
It is understood that the limit $n\to 0$ is to be taken at the end 
of any calculation.  In terms of replicas, the expectation value of 
this field $\langle\wh{\Omega}(x)\rangle$ is proportional to 
\begin{equation}
\left\langle
J^{-1}\sum\nolimits_{j=1}^{J}
\prod\nolimits_{\alpha=0}^{n}\delta
\big(
{\bf x}^{\alpha}-{\bf R}_{j}^{\alpha}
\big)
\right\rangle
\label{eq:PDinterpret}
\end{equation}
and its Fourier transform 
$\op(k)=\int\! dx\,\exp(ik\cdot x)\,\wh{\Omega}(x)$, 
has the form 
\begin{equation}
\Omega(k)=
\left\langle
J^{-1}\sum\nolimits_{j=1}^{J}
\exp i
\sum\nolimits_{\alpha=0}^{n}
{\bf k}^{\alpha}
\cdot 
{\bf R}_{j}^{\alpha}
\right\rangle.
\end{equation}
The field $\wh{\Omega}(x)$ fluctuates subject to the demand, mentioned
above, that the critical freedoms are only the corresponding Fourier
amplitudes $\op(k)$ for which at least two of the $D$-component
entries in the argument 
$k\equiv\{{\bf k}^{0},{\bf k}^{1},\ldots,{\bf k}^{n}\}$ 
are nonzero.  Both a semi-microscopic approach and
arguments based on symmetries and length-scales yield a Landau-Wilson
effective Hamiltonian governing the fluctuations of this field, which
is invariant under independent translations of the replicas but whose
precise structure we shall discuss later.  For now, let us just mention
that the corresponding expectation value of this field is the order
parameter~(\ref{eq:phyop}) with $\nu=1+n$: it becomes nonzero in the
amorphous solid state and, in doing so, realizes the pattern of
spontaneous symmetry breaking described above.  Invariance under
independent translations of the replicas breaks down to invariance
under the subgroup of common translations of the replicas.

\section{Goldstone fluctuations: 
         Structure and identification}
\label{sec:GoldstoneStructure}
\subsection{Formal construction of Goldstone fluctuations}
\label{sec:formal}
In the amorphous solid state, one of the symmetry-related family 
of classical values of the order parameter has the form
\begin{equation}
\op(k)=
\delta_{{\bf k}_{\rm tot},{\bf 0}}\,
\cw(k_{\rtr})=
\int_{\cal V} \frac{d{\bf x}_{\rm cm}}{V}\,
e^{i{\bf k}_{\rm tot}\cdot{\bf x}_{\rm cm}}\,
\cw(k_{\rtr}), 
\label{eq:classicalstate}
\end{equation}
in which $\cw$ is real and depends only on the magnitude of $k_{\rtr}$.
Some geometry is needed to define the variables in this formula.
We introduce a complete orthonormal basis set in replica space
$\{\ers^{\alpha}\}_{\alpha=0}^{n}$, in terms of which vectors $k$
are expressed as
\begin{equation}
k=\sum\nolimits_{\alpha=0}^{n}
{\bf k}^{\alpha}\,\ers^{\alpha}.
\label{eq:basisvector}
\end{equation}
We also introduce the {\it replica body-diagonal\/} unit vector
\begin{equation}
\ers\equiv\frac{1}{\sqrt{1+n}}
\sum\nolimits_{\alpha=0}^{n}\ers^{\alpha},
\label{eq:bodydiagonal}
\end{equation}
relative to which we may decompose vectors $k$ into
longitudinal ($\rlo$) and transverse ($\rtr$) components: 
\begin{equation}
k=k_{\rlo}+k_{\rtr},\quad
k_{\rlo}\equiv  (k\cdot\ers)\,\ers,\quad
k_{\rtr}\equiv   k-(k\cdot\ers)\,\ers. 
\label{eq:tldecomp}
\end{equation}
We find it convenient to parametrize the longitudinal 
components of position and wave vectors in the 
following distinct ways: 
\begin{subequations}
\begin{eqnarray}
x_{\rlo}&=&(1+n)^{ 1/2}\,{\bf x}_{\rm cm }\,\ers,
\quad\!\!
{\bf x}_{\rm cm }\equiv\frac{1}{1+n}
\sum_{\alpha=0}^{n}{\bf x}^{\alpha},\\
k_{\rlo}&=&(1+n)^{-1/2}\,{\bf k}_{\rm tot}\,\ers,
\quad\!\!
{\bf k}_{\rm tot}\equiv\sum_{\alpha=0}^{n}{\bf k}^{\alpha}.
\label{eq:TotalMom}
\end{eqnarray}
\end{subequations} 
Then ${\bf x}_{\rm cm }$ and ${\bf k}_{\rm tot}$ are, respectively, 
the analogs of the following conjugate pair of vectors: 
the center-of-mass position and the total momentum.  
With them  one then has, e.g.,
$k_{\rlo}\cdot x_{\rlo}={\bf k}_{\rm tot}\cdot{\bf x}_{\rm cm}$.

 \begin{figure}
 \centerline{\psfig{figure=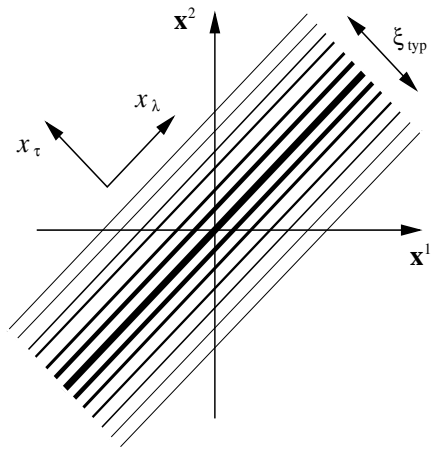,width=6cm,height=6cm,angle=0}}
 \vspace{-0.2cm} 
 \caption{A classical state in replicated real space: a hill in $x$ space with 
 its ridge aligned in the $x_{\rlo}$ direction and passing through the origin.  
 The thickness of the lines is intended to suggest the amplitude of the order 
 parameter, the thicker the line the larger the amplitude.  Symmetry-related 
 classical states follow from rigid displacements of the hill perpendicular 
 to the ridge, i.e., in the $x_{\rtr}$ direction.}
 \label{fig:hill}
 \end{figure}
The variables just introduced exhibit the structure of a classical 
state Eq.~(\ref{eq:classicalstate}) as a rectilinear {\it hill\/} 
in $x$-space, with contours of constant height oriented along 
$x_{\rlo}$, as shown in Fig.~\ref{fig:hill}.  The peak height of 
the ridge determines the fraction of localized particles; the 
decay of the height in the direction $x_{\rtr}$ determines the 
distribution of localization lengths.  The width of the hill 
corresponds to the typical value of the localization length.  
Symmetry-related classical states are generated from 
Eq.~(\ref{eq:classicalstate}) by translating the hill rigidly, 
perpendicular to the ridge-line (i.e.~parallel to $x_{\rtr}$). 
Such a transformation corresponds to relative (but not common) 
translations of the replicas.

 \begin{figure}
 \centerline{\psfig{figure=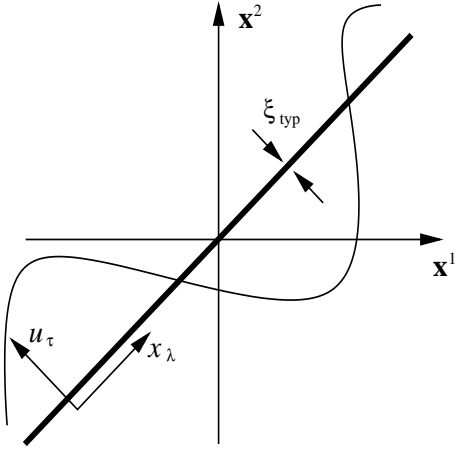,width=6cm,height=6cm,angle=0}}
 \vspace{-0.2cm}
 \caption{Goldstone-distorted  state in replicated real space: 
 the hill is displaced perpendicular to the ridge to an extent 
 $u_{\rtr}$ that varies with position $x_{\rlo}$ along the ridge. 
 Note that the scale of this figure is much larger than that 
 used for Fig.~\ref{fig:hill}: the thick line lies along the 
 ridge of the classical state, but now it is the {\it width\/} 
 of the line that indicates the width of the hill $\xi_{\rm typ}$.  
 The Goldstone-type fluctuations occur on wave-lengths longer than 
 $\xi_{\rm typ}$.}
 \label{fig:ripple}
 \end{figure}
This pattern of symmetry breaking suggests that the Goldstone 
excitations of a classical state are constructed from it via 
$x_{\rlo}$ (or equivalently ${\bf x}_{\rm cm}$) -dependent 
translations of the hill in the $x_{\rtr}$ direction, i.e., 
ripples of the hill and its ridge.  In two equivalent 
realizations, this gives
\begin{subequations}
\begin{eqnarray}
\label{eq:GDk}
V\op(k)\!\!&=&\!\!
\int_{\cal V} d{\bf x}_{\rm cm}\,
e^{i{\bf k}_{\rm tot}\cdot{\bf x_{\rm cm}}+
   ik_{\rtr}\cdot u_{\rtr}({\bf x}_{\rm cm})}\,
\cw(k_{\rtr}),\\
\label{eq:GDx}
V\wh{\op}(x)\!\!&=&\!\!
{\wh\cw}\left(x_{\rtr}-u_{\rtr}({\bf x}_{\rm cm})\right).
\end{eqnarray}
\end{subequations}
Note that we are choosing to define Fourier transforms as follows: 
\begin{subequations}
\begin{eqnarray}
\wh{A}(x)&=&
\int\dbar k_{\rtr}\,\,\dbar{\bf k}_{\rm tot}\,
e^{-ik_{\rtr}\cdot x_{\rtr}}\,
e^{-i{\bf k}_{\rm tot}\cdot{\bf x}_{\rm cm}}\, 
{A}(k), 
\\
{A}(k)&=&
\int dx_{\rtr}\,d{\bf x}_{\rm cm}\,
e^{ik_{\rtr}\cdot x_{\rtr}}\,
e^{i{\bf k}_{\rm tot}\cdot{\bf x}_{\rm cm}}\, 
\wh{A}(x); 
\end{eqnarray}
also note that 
\begin{equation}
{\wh\cw}(x_{\rtr})\equiv
\int\dbar k_{\rtr}\,
e^{-ik_{\rtr}\cdot x_{\rtr}}\,
{\cw}(k_{\rtr}).
\end{equation}
\end{subequations}
Here and elsewhere, bars indicate division by factors of $2\pi$;  
on integration measures there is one such division for each 
variable of integration. 
The details of the excitation are encoded in the replica-transverse 
field $u_{\rtr}({\bf x}_{\rm cm})$, an $nD$-component field that 
depends on the replica-longitudinal position ${\bf x}_{\rm cm}$; these 
are the {\it Goldstone bosons\/}, or phonon excitations, of the 
amorphous solid state.  Provided the Fourier content of the Goldstone 
field $u_{\rtr}({\bf x}_{\rm cm})$ occurs at wave-lengths long compared 
with the hill width (i.e.~the typical localization length)---a point 
that we shall amplify below---these exhaust the spectrum of 
low energy excitations. 

The Goldstone excitations that we have just constructed are analogs 
of the capillary excitations of the interface between coexisting liquid 
and gas states; see Ref.~\cite{Wallace1982} for a review.  In the liquid-gas 
context they similarly accompany a spontaneous breaking of translational 
symmetry associated with the choice of interface location.

What about other excitations of the broken-symmetry state?  
The complete spectrum of exciations is accounted for by 
decorating the Goldstone-type parametrization~(\ref{eq:GDk}) 
with additional freedoms $\mw(k_{\rtr})$, so that the field 
$\op(k)$ is expressed as  
\begin{equation}
V\op(k)=\int_{\cal V} d{\bf x}_{\rm cm}\,
e^{i{\bf k}_{\rm tot}\cdot{\bf x}_{\rm cm}+
   ik_{\rtr}\cdot u_{\rtr}({\bf x}_{\rm cm})}
   \Big(\cw(k_{\rtr})+\mw(k)\Big), 
\label{eq:FullSpectrum}
\end{equation}
where $\mw$ is a real-valued field that depends on $k_{\rtr}$, 
suitably constrained to be independent of the Goldstone excitations. 

The structure of the Goldstone excitations is in accordance with the 
linear stability analysis of the classical broken-symmetry state, due 
to Castillo et al.~\cite{Castillo+Goldbart+Zippelius1999}, which 
identified a family of linearly additive Goldstone-type normal modes 
of excitation indexed by ${\bf k}_{\rm tot}$: 
\begin{equation}
V\op(k)=
\int_{\cal V} d{\bf x}_{\rm cm}\,
e^{i{\bf k}_{\rm tot}\cdot{\bf x}_{\rm cm}}\,\cw(k_{\rtr}) 
+ik_{\rtr}\cdot v_{\rtr}({\bf k}_{\rm tot})\,\cw(k_{\rtr}),
\label{eq:FlAdditive}
\end{equation}
with arbitrary replica-transverse amplitude $v_{\rtr}({\bf k}_{\rm tot})$. 
That this linear glimpse of the Goldstone-type excitations is in accord 
with the nonlinear view focused on in the present Paper follows by 
expanding Eq.~(\ref{eq:FullSpectrum}) to linear order in the Goldstone 
fields $u_{\rtr}$ and omitting the non-Goldstone fields $\mw$, 
thus arriving at 
\begin{eqnarray}
&&V\op(k)
\approx
\int_{\cal V} d{\bf x}_{\rm cm}\,
 e^{i{\bf k}_{\rm tot}\cdot{\bf x}_{\rm cm}}\,\cw(k_{\rtr})
\nonumber\\ 
&&\quad+
ik_{\rtr}\cdot
\int_{\cal V} d{\bf x}_{\rm cm}\,
e^{i{\bf k}_{\rm tot}\cdot{\bf x}_{\rm cm}}\,
u_{\rtr}({\bf x}_{\rm cm})\,
\cw(k_{\rtr}), 
\label{eq:FlExpander}
\end{eqnarray}
which shows that $v_{\rtr}({\bf k}_{\rm tot})$ is the 
Fourier transform of $u_{\rtr}({\bf x}_{\rm cm})$.  
The reality of $\op(x)$ ensures that $\op(-k)=\op(k)^{\ast}$ and, 
via Eq.~(\ref{eq:FlAdditive}), that
$v_{\rtr}(-{\bf k}_{\rm tot})=v_{\rtr}({\bf k}_{\rm tot})^{\ast}$. 
Via Eq.~(\ref{eq:FlExpander}), this in turn ensures the reality of 
$u_{\rtr}({\bf x}_{\rm cm})$.

\begin{figure}
 \centerline{\psfig{figure=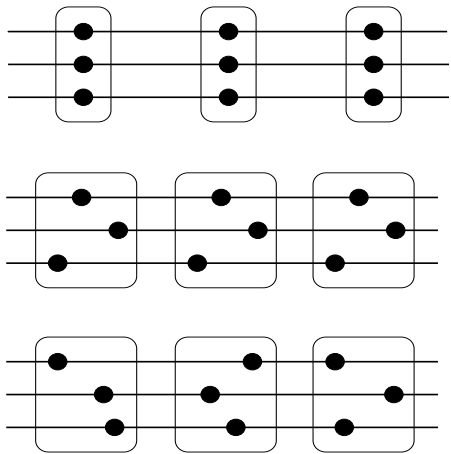,width=6cm,height=6cm,angle=0}}
\vspace{-0.2cm}
\caption{{\it Molecular bound state\/} view of the classical 
and Goldstone-distorted states in replicated real space.  
The full circles within a border represent the replicas of a given 
monomer location.  Repetitions of them along a row indicate that 
the center of mass of the bound states is distributed homogeneously. 
Upper bars: 
one classical state; the probability density is peaked at the 
shown configurations, in a manner independent of the location of 
the center of mass of the bound state of the replicated particles. 
Middle bars: 
another classical state, obtained by the former one via a relative 
translation of the replicas that does not vary with the location 
of the center of mass. 
Lower bars: 
a Goldstone-distorted state; the the probability density is 
peaked in a manner that varies with the location of 
the center of mass.} 
\label{fig:molecule}
\end{figure}
Some insight into the structure of the Goldstone excitations, which 
induce deformations of the classical state, is obtained from its 
replicated real-space version.  Consider the interpretation of 
$\wh{\op}(x)$ as a quantity proportional to the probability 
density for the positions of the $1+n$ replicas of a particle to 
have the values $\{{\bf x}^{\alpha}\}_{\alpha=0}^{n}$; 
see Eq.~(\ref{eq:PDinterpret}).  Then classical states, 
Eq.~(\ref{eq:GDx}) at constant $u_{\rtr}$, are ones that describe 
a translationally invariant {\it bound states\/}: $\wh{\op}(x)$ does 
not depend on the {\it mean\/} location of the replicas 
${\bf x}_{\rm cm}$, but does depend on their {\it relative\/} 
locations, through $x_{\rtr}$, and decays the more 
the replicas are separated.  This point is exemplified 
by the particular form given in Eq.~(\ref{eq:mfstate}).  
Now, the Goldstone-distorted state, Eq.~(\ref{eq:GDx}) with $u_{\rtr}$ 
varying with ${\bf x}_{\rm cm}$, also describes bound states of the 
replicas, but ones in which the dependence of the probability density 
on relative locations varies with ${\bf x}_{\rm cm}$.  The particular 
form~(\ref{eq:mfstate}), which gives  
\begin{eqnarray}
&&V\wh\op(x)=Q\int_{0}^{\infty}\!\!\! 
d\xi^{2}\,\dsl(\xi^{2})\,
\left(2\pi\xi^{2}\right)^{-nD/2}
\nonumber\\
&&\qquad\qquad\quad\times
\exp\left(-
\vert{x_{\rtr}-u_{\rtr}}({\bf x}_{\rm cm})\vert^{2}/2\xi^{2}
\right), 
\label{eq:RSgold}
\end{eqnarray}
exemplifies this point; in particular, one sees that the most probable 
value $u_{\rtr}$ of the {\it relative\/} locations $x_{\rtr}$ now 
depends on the {\it center-of-mass\/} location ${\bf x}_{\rm cm}$.  
These remarks are amplified in Fig.~\ref{fig:molecule}.

Returning to the issue of the structure and properties of the 
Goldstone-type excitations of the amorphous solid state, 
recall that the critical Fourier 
amplitudes of the field are those that reside in the
higher-replica sector [i.e.~HRS, for which at least two $D$-vector elements of
the argument of $\op(k)$ are nonzero].  Do the proposed Goldstone distortions
of the classical state excite the lower-replica sectors?  If so, they would be
suppressed by interactions, such as particle repulsion, that tend to preserve
homogeneity.  To see that they do not, let us examine the Goldstone-distorted
state in the zero- (i.e.~$k=0$) and one- (i.e.~$k={\bf q}\,\ers^{\alpha}$ with
${\bf q}\ne{\bf 0}$) replica sectors.  In the former, one readily sees from
Eq.~(\ref{eq:GDk}) that it has its undistorted value.  A straightforward
calculation shows that in the latter sector the distorted order parameter
contains the factor:
\begin{equation}
\int_{\cal V}
\frac{d{\bf x}_{\rm cm}}{V}\,
\exp\big(
 i{\bf q}\cdot{\bf x}_{\rm cm}
+i{\bf q}\cdot{\bf u}^{\alpha}({\bf x}_{\rm cm})
\big),
\label{eq:integrateconstraint}
\end{equation}
where ${\bf u}^{\alpha}=u_{\rtr}\cdot\ers^{\alpha}$.
By introducing the transformation
${\bf x}_{\rm cm}\to
 {\bf x}_{\rm cm}^{\prime}\equiv
 {\bf x}_{\rm cm}+{\bf u}^{\alpha}({\bf x}_{\rm cm})$
we see that the Goldstone-distorted state remains zero in the 
one-replica sector, provided all $D$-vector elements ${\bf u}^{\alpha}$ 
of the Goldstone field $u_{\rtr}({\bf x}_{\rm cm})$ obey the condition
\begin{equation}
\big\vert\det
\left(\delta_{dd^{\prime}}^{\phantom\alpha}
+\partial_{d}^{\phantom\alpha}u_{d^{\prime}}^{\alpha}
\right)\big\vert=1,
\label{eq:differentialconstraint}
\end{equation}
which, for small amplitude distortions, reduces to
$\partial_{d}^{\phantom\alpha}u_{d}^{\alpha}=0$. 
That this condition corresponds to incompressibility
(i.e.~{\it pure shear\/}) will be established 
in Sec.~\ref{sec:GFwithoutReplicas}.
(Here and elsewhere, summations from $1$ to $D$ are implied over 
repeated Cartesian indices, such as $d$ and $d^{\prime}$.)\thinspace\ 
This is as it should be: to make density fluctuations there must be some
compression amongst the elements ${\bf u}^{\alpha}$ of $u_{\rtr}$.

\subsection{Identifying the Goldstone fluctuations as local displacements}
\label{sec:GFwithoutReplicas}
In this subsection our aim is to establish the connection between
the Goldstone fields and the displacement fields of conventional 
elasticity theory~\cite{LandauLifshitz,KleinertGFS}.  
{\it Inter alia\/}, this identifies the stiffness associated with 
the Goldstone fluctuations as the elastic modulus governing shear 
deformations of the amorphous solid. 
As we shall see, the discussion is general enough to apply to  
\emph{any} amorphous solid that breaks translational symmetry
microscopically but preserves it macroscopically, in regimes 
where the constituent particles are strongly localized in position. 

Recall the amorphous solid order parameter, Eq.~(\ref{eq:phyop}),  
\begin{equation}
\Big[
J^{-1}\sum\nolimits_{j=1}^{J}
\langle{\rm e}^{i{\bf k}^{1}\cdot{\bf R}_{j}}\rangle
\langle{\rm e}^{i{\bf k}^{2}\cdot{\bf R}_{j}}\rangle
\cdots
\langle{\rm e}^{i{\bf k}^{\nu}\cdot{\bf R}_{j}}\rangle
\Big], 
\nonumber
\end{equation}
and focus on a single element 
$\langle{\rm e}^{i{\bf k}\cdot{\bf R}_{j}}\rangle$. 
It is convenient to consider the element in the form  
\begin{equation}
{\rm e}^{i{\bf k}\cdot{\bf r}_{j}}\,
\cloud_{j}({\bf k}), 
\end{equation}
where ${\bf r}_{j}$ is the mean of the position of particle $j$,  
and $\cloud$ describes fluctuations about the mean. 
Now consider the impact of a long wave-length shear displacement, 
encoded in the field $\sdcon({\bf x})$, which deforms the 
probability density associated with the position ${\bf R}_{j}$ 
of particle $j$.  By assuming that the typical localization length 
is much smaller than the length-scale associated with variations of 
the deformation we are able to retain only the {\it rigid\/} 
displacement of the probability density for ${\bf R}_{j}$ and 
neglect any deformation of its {\it shape\/}.  Thus, under the 
displacement the element is deformed as 
\begin{equation}
\langle{\rm e}^{i{\bf k}\cdot{\bf R}_{j}}\rangle=
{\rm e}^{i{\bf k}\cdot{\bf r}_{j}}\,
\cloud_{j}({\bf k}) 
\longrightarrow
{\rm e}^{i{\bf k}\cdot\left({\bf r}_{j}+\sdcon({\bf r}_{j})\right)}\,
\cloud_{j}({\bf k}), 
\end{equation}
Inserting such deformations into the order parameter, with an 
independent displacement field for each element, gives the 
distorted form 
\begin{eqnarray}
&&
\Big[\Jinver\sum_{j=1}^{J}
{\rm e}^{i\sum_{a=1}^{\nu}{\bf k}^{a}\cdot{\bf r}_{j}}\,
{\rm e}^{i\sum_{a=1}^{\nu}{\bf k}^{a}\cdot\sdcon^{a}({\bf r}_{j})}\,
\cloud_{j}({\bf k}^{1})\cdots\cloud_{j}({\bf k}^{\nu})\Big]
\nonumber\\
&&=
\int\frac{d{\bf r}}{V}\,
{\rm e}^{i\sum_{a=1}^{\nu}{\bf k}^{a}\cdot
        \left({\bf r}+\sdcon^{a}({\bf r})\right)}\,
\Big[\cloud_{j}({\bf k}^{1})\cdots\cloud_{j}({\bf k}^{\nu})\Big], 
\end{eqnarray}
where we have arrived at the second form by noting that, in the amorphous 
solid state, the mean position of any particle is distributed 
homogeneously.  Next, we decompose the collection of displacement 
fields $\{\sdcon^{a}\}$ into longitudinal and transverse 
parts, according to the geometrical prescription given in 
Sec.~\ref{sec:formal}, but keeping in mind the fact that there are now 
$\nu$ copies (rather than $1+n$).  The essential point is that the 
longitudinal part of $\{\sdcon^{a}\}$, which corresponds to {\it common\/}
deformations of the elements, does not generate a new value of the order 
parameter, in contrast with the transverse part (which generates 
{\it relative\/} deformations).  Therefore, the {\it physical\/} 
displacements are the transverse part of $\{\sdcon^{a}\}$.
To see this point, consider the special situation in which the transverse  
part of $\{\sdcon^{a}\}$ {\it is\/} position dependent but the longitudinal 
part {\it is not\/}.  In this case, ($V$ times) the order parameter becomes 
\begin{equation}
\int{d{\bf r}}\,
{\rm e}^{i\sum_{a=1}^{\nu}{\bf k}^{a}\cdot{\bf r}}\,
{\rm e}^{
ik_{\rtr}\cdot v_{\rtr}({\bf r})+
i{\bf k}_{\rm tot}\cdot\sdcon_{\rm cm}}\,
\Big[\cloud_{j}({\bf k}^{1})\cdots\cloud_{j}({\bf k}^{\nu})\Big].
\label{eq:goal}
\end{equation} 
As things stand, in the presence of $v_{\rtr}({\bf r})$, a 
longitudinal part $\sdcon_{\rm cm}$ would have the effect of 
producing a new value of the order parameter.  As, on physical 
grounds, we expect that it should not, we see that {\it physical\/} 
displacements are purely transverse.   This is a consequence of the 
fact that, for the long wave-length displacements under consideration, 
the deformed amorphous solid state continues to preserve translational 
invariance macroscopically.  This argument for the absence of the 
longitudinal part of the displacement field continues to hold when 
it has position-dependence: if, when constant, it does not generate 
a new state degenerate with the old one then, when varying, it should 
not generate a low-energy deformation. 

Thus we have realized the goal of this section: by comparing 
Eqs.~(\ref{eq:GDk}) and (\ref{eq:goal}) we see that the 
formally-constructed Goldstone fields $u_{\rtr}$ are in fact the 
physical displacement fields $v_{\rtr}$.  Actually, there is one 
further point to address, concerning the argument of the final 
factor in Eq.~(\ref{eq:goal}), viz., 
$\big[\,\cloud_{j}({\bf k}^{1})\cdots\cloud_{j}({\bf k}^{\nu})\,\big]$.  
Apparently, there is dependence on the full set of wave vectors 
$\{{\bf k}^{1},\ldots,{\bf k}^{\nu}\}$, whereas Eq.~(\ref{eq:GDk}) 
indicates dependence only on the transverse part of this collection. 
The resolution of this apparent discrepancy lies in the observation 
that, for position-independent $v_{\rtr}$, Eq.~(\ref{eq:goal}) must 
revert to a classical state, for which the final factor does not 
depend on ${\bf k}_{\rm tot}$.  As the final factor consists of 
probability clouds, which we are assuming to be undeformed by the 
displacements, this factor continues to be independent of 
${\bf k}_{\rm tot}$ in Eq.~(\ref{eq:goal}).

\section{Energetics of Goldstone fluctuations; elastic free energy}
\label{sec:GoldstoneEnergetics}
The simplest Landau-Wilson effective hamiltonian controlling the order
parameter $\op$ has the form
\begin{eqnarray}
&&{\cal S}_{\op}=
V\den\sum_{k\in {\hrs}}
\left(-a\tau+
\smhalf
\lsz^{2}\,
k\cdot k\right)
\op(k)\,\op(-k)
\label{eq:Landaueffective}
\\
&&\quad
-V\den\,g
\sum_{k_{1},k_{2},k_{3}\in\hrs}\!\!\!\!
\delta_{k_{1}+k_{2}+k_{3},0}\,\,
\op(k_{1})\,\op(k_{2})\,\op(k_{3}),
\nonumber
\end{eqnarray}
where $\den$ is the number of entities 
being constrained per unit volume,
and $\{a\tau,\lsz,g\}$ are, respectively,
the control parameter for the density of constraints,
the linear size of the underlying objects being linked,
and the nonlinear coupling constant controlling the strength with
which the $\op$ fluctuations interact.
$\hrs$ indicates that only wave vectors in the higher-replica sector
are to be included in the summations.  A semi-microscopic model of
vulcanized macromolecular matter yields
$a=1/2$,
$\tau=(\newmu^{2}-\newmu_{\rm c}^{2})/\newmu_{\rm c}^{2}$,
$\lsz^{2}=L\lpc/2D$ and
$g=1/6$,
where $\newmu$ controls the mean number of constraints
(and has critical value $\newmu_{\rm c}=1$),
$\lpc$ is the persistence length of the macromolecules,
and $L/\lpc$ is the number of segments per macromolecule.

In terms of ${\cal S}_{\op}$, the replica partition function 
$\rpf_{1+n}$ is given integrating over the critical modes of $\op$:
\begin{equation}
\rpf_{1+n}\sim\int\fim\op\,e^{-{\cal S}_{\op}}.
\label{eq:replicapartition}
\end{equation}
Then, according to the replica scheme of Deam and 
Edwards~\cite{Deam+Edwards1975},
the disorder-averaged free energy $F$ is given by
\begin{equation}
-\frac{F}{T}=
\lim_{n\to 0}\,\,\frac{\rpf_{1+n}-\rpf_{1}}{n\rpf_{1}}
=\lim_{n\to 0}\,\,\frac{\partial}{\partial n}\ln\rpf_{1+n},
\end{equation}
where $T$ is the temperature.

It was shown in Ref.~\cite{Castillo+Goldbart+Zippelius1994} (see also 
Refs.~\cite{PengCastillo1998,GoldbartTrieste2000,GoldbartAdvPhy1996}) 
that making ${\cal S}_{\op}$ stationary with respect to $\op$ 
results in the classical state~(\ref{eq:classicalstate}), with
\begin{subequations}
\label{eq:CSelements}
\begin{eqnarray}
\cw(k_{\rtr})&=&
Q\int_{0}^{\infty}d\xi^{2}\,\dsl(\xi^{2})\,
e^{-\xi^{2}k_{\rtr}^{2}/2},
\label{eq:mfstate}
\\
\label{eq:gelfraction}
Q&=&2a\tau/3g,
\\
\noalign{\medskip}
\dsl(\xi^{2})&=&
\big(\lsz^{2}/a\tau\xi^{4}\big)\,
\pi\big(\lsz^{2}/a\tau\xi^{2}\big),
\label{eq:distribution}
\end{eqnarray}
\end{subequations}
where $\pi(\theta)$ is the universal classical scaling function 
discussed in 
Refs.~\cite{Castillo+Goldbart+Zippelius1994,PengCastillo1998}.  
Distorting the classical state, 
via Eq.~(\ref{eq:GDk}) or (\ref{eq:GDx}), inserting the resulting 
state into ${\cal S}_{\op}$ and computing the {\it increase\/}, 
${\cal S}_{u}$, in ${\cal S}_{\op}$ due to the distortion $u_{\rtr}$
(i.e.~the elastic free energy) gives the following contribution, 
which arises solely from the quadratic \lq\lq gradient\rq\rq\ term 
in ${\cal S}_{\op}$:
\begin{subequations}
\label{eq:elasticcouple}
\begin{eqnarray}
\label{eq:elasticenergy}
{\cal S}_{u}&=&
\frac{\stiff_{n}}{2T}
\int_{\cal V} d{\bf x}\,
\left(
\partial_{\bf x}u_{\rtr}\cdot
\partial_{\bf x}u_{\rtr}\right),
\\
\label{eq:replicastiffness}
\stiff_{n}&\equiv&
\frac{T\,\den}{(1+n)^{1+\frac{D}{2}}}
\int 
V^{n}\,
\dbar k_{\rtr}\,
\frac{\lsz^{2}\,k_{\rtr}^{2}}{nD}\,
\cw(k_{\rtr})^{2},
\end{eqnarray}
\end{subequations}
in the former of which there are scalar products 
over both the $nD$ independent components of $u_{\rtr}$ and the $D$ 
components of ${\bf x}$.  The derivation of this elastic free energy 
is given in App.~\ref{sec:LWHtoEFE}.
As we have discussed in Sec.~\ref{sec:GFwithoutReplicas}, 
and shall revisit in Sec.~\ref{sec:ShearModulus}, 
$\stiff_{0}$ is the elastic shear modulus.
By using the specific classical form for $\cw$, Eq.~(\ref{eq:mfstate}), 
and passing to the replica limit, $n\to 0$, we obtain
\begin{subequations}
\label{eq:elasticstiffness}
\begin{eqnarray}
\stiff_{0}
&=&
T\den\,Q^{2}
\int
d\xi^{2}\,\dsl(\xi^{2})\,
d\xi^{\prime 2}\,\dsl(\xi^{\prime 2})\,
\frac{\lsz^{2}}{\xi^{2}+\xi^{\prime 2}}
\\
&=&
T\den\,\tau^{3}\,
\frac{4 a^{2}}{9 g^{2}}
\int d\theta\,d\theta^{\prime}\,
\frac{\pi(\theta)\,\pi(\theta^{\prime})}{\theta^{-1}+\theta^{\prime -1}}
\end{eqnarray}
which, for the case of the semi-microscopic parameters stated shortly 
after Eq.~(\ref{eq:Landaueffective}), becomes
\begin{equation}
\stiff_{0}=
2\,T\,\den\,\tau^{3}
\int d\theta\,d\theta^{\prime}\,
\frac{\pi(\theta)\,\pi(\theta^{\prime})}{\theta^{-1}+\theta^{\prime -1}}\,.
\label{eq:shearscale}
\end{equation}
\end{subequations}
The main technical steps of this derivation are given in  
App.~\ref{sec:ClassShear}.  Thus, we  have arrived at the effective 
free energy controlling elastic deformations, in the harmonic 
approximation.  It is consistent with the result obtained in 
Refs.~\cite{Castillo+Goldbart1998-2000}, in which the free energy cost 
of imposing a macroscopic shear deformation of the sample was determined.  

\section{Identification of the shear modulus: 
         Macroscopic view}
\label{sec:ShearModulus}
In Sec.~\ref{sec:GFwithoutReplicas}, we have explained why the 
Goldstone-type fluctuations are identified with local displacements 
of the amorphous solid by considering the impact of these fluctuations 
at a semi-microsopic level.  In the present section we again address 
this identification, but now from a more macroscopic perspective, by 
coupling the Goldstone fields to a force-density field.

Accounting solely for the Goldstone-type fluctuations 
[i.e.~ignoring the field $\mw$ in the parametrization of 
the field $\op$ in Eq.~(\ref{eq:FullSpectrum})], we approximate 
the replica partition function~(\ref{eq:replicapartition}) as
\begin{eqnarray}
&&\!\!\!\!\rpf_{1+n}[f_{\rtr}]\sim
e^{-{\cal S}_{\op,{\rm cl}}}
\!\!\int\!\!\fim u_{\rtr}
\exp\Big(\!
-\frac{\stiff_{0}}{2T}
\!\int_{\cal V}\! d{\bf x}\,
\partial_{\bf x}u_{\rtr}\!\cdot
\partial_{\bf x}u_{\rtr}
\nonumber
\\
&&
\qquad\qquad\qquad\qquad
+\frac{1}{T}\int_{\cal V} d{\bf x}\,
f_{\rtr}({\bf x})\cdot u_{\rtr}({\bf x})
\Big),
\label{eq:gausspartition}
\end{eqnarray}
in which ${\cal S}_{\op,{\rm cl}}$ is the effective 
free energy~(\ref{eq:Landaueffective}) evaluated
in the classical state~(\ref{eq:classicalstate}), 
and $\fim u_{\rtr}$ indicates functional integration over 
replicated displacement fields 
$\{{\bf u}^{\alpha}({\bf x})\}$
subject to the following conditions: 
$u$ is replica-transverse 
[cf.~Eq.~(\ref{eq:tldecomp})]; 
the $D$-vector elements it contains are pure shear 
[cf.~Eq.~(\ref{eq:differentialconstraint})]; 
and the Fourier content is restricted to wave-lengths 
longer than a short-distance cut-off $\coshort$ 
(which is we take to be on the order of the typical localization length)
but shorter than a long-distance cut-off $\colong$ 
(which is commonly on the order of the linear size of the sample). 
The reason for the restriction to wave-lengths longer than the typical 
localization length is that by restricting our attention to the Goldstone 
sector of fluctuations we are omitting the effects of massive flutuations. 
To be consistent, we should also omit the effects of Goldstone-type 
fluctuations with wave-lengths sufficiently short that their energy 
scale is comparable to or larger than the scale for the (omitted) least 
massive fluctuations.  The appropriate criterion is that the short-distance 
cut-off be taken to be on the order of the typical localization length.  
This can be appreciated pictorially from Figs.~\ref{fig:hill} and 
\ref{fig:ripple}: Goldstone-type fluctuations having wave-lengths smaller 
that the hill width are omitted.  Note that we have ignored the Jacobian 
factor connected with the change of functional integration variable from 
$\op$ to $u$. 

In order to reconfirm the identification of the shear modulus, we have, 
in Eq.~(\ref{eq:gausspartition}), coupled the displacement field $u$ 
linearly to a replicated {\it force density field\/} $f$ though a term 
$-\frac{1}{T}\int d{\bf x}\,f({\bf x})\cdot u({\bf x})$; 
because $u_{\rlo}$ is zero, only the transverse term, 
$-\frac{1}{T}\int d{\bf x}\,
f_{\rtr}({\bf x})\cdot u_{\rtr}({\bf x})$, remains.  
As for $f$ itself, it is taken to have $D$-vector elements that vanish 
in the zeroth replica and are identically equal to ${\bf f}$ in the 
remaining replicas.  This reflects the fact that the force density 
is envisaged as being applied {\it subsequent\/} to the cross-linking 
process, and therefore does not feature in the replica that generates 
the cross-link distribution, but is repeated in the {\it thermodynamic\/} 
replicas (by which we mean the replicas that generate the logarithm of 
the partition function, not the disorder distribution).  Thus, one has
\begin{eqnarray}
f_{\rtr}\cdot f_{\rtr}
&=&f\cdot f-f_{\rlo}\cdot f_{\rlo}
=n\,{\bf f}\cdot{\bf f}-(f\cdot\ers)^{2}
\nonumber\\ 
\noalign{\medskip}
&=&\frac{n}{1+n}\,{\bf f}\cdot{\bf f}
\mathrel{\mathop{\longrightarrow}^{n\to 0}}n\,{\bf f}\cdot{\bf f}.
\end{eqnarray}

\begin{widetext}
The integration over $u$ in Eq.~(\ref{eq:gausspartition}) is gaussian,
and thus straightforward, requiring only the elastic correlator
$\langle u_{d}({\bf y})\,u_{d^{\prime}}({\bf y}^{\prime})\rangle$,
which is given in terms of the elastic Green function
${\cal G}_{dd^{\prime}}({\bf y}-{\bf y}^{\prime})$:
\begin{subequations}
\begin{eqnarray}
\langle u_{d}({\bf y})\,
        u_{d^{\prime}}({\bf y}^{\prime})
\rangle
&\equiv&
\frac{
\int\fim{\bf u}\,
\exp\left(
-\frac{\stiff_{0}}{2T}
\int_{\cal V}d{\bf x}\,
\partial_{\bf x}{\bf u}\cdot
\partial_{\bf x}{\bf u}
    \right)\,
u_{d}({\bf y})\,
u_{d^{\prime}}({\bf y}^{\prime})}
{\int\fim{\bf u}\,
\exp\left(
-\frac{\stiff_{0}}{2T}
\int_{\cal V}d{\bf x}\,
\partial_{\bf x}{\bf u}\cdot
\partial_{\bf x}{\bf u}
    \right)}=
\frac{T}{\stiff_{0}}\,\,
{\cal G}_{dd^{\prime}}({\bf y}-{\bf y}^{\prime}), 
\label{eq:uuviawick}
\\
{\cal G}_{dd^{\prime}}({\bf y})
&=&
\int_{2\pi/\colong}^{2\pi/\coshort}
\dbar{\bf k}\,
e^{-i{\bf k}\cdot{\bf y}}\,
{\cal G}_{dd^{\prime}}({\bf k}),
        \qquad\qquad
{\cal G}_{dd^{\prime}}({\bf k})
= 
\big(k^{2}\,\delta_{dd^{\prime}}-k_{d}\,k_{d^{\prime}}\big)/k^{4}, 
\label{eq:greenKS}
\end{eqnarray}
\end{subequations}
where $\colong$ and $\coshort$ are, respectively, the long- and 
short-distance cut-offs on the wave-vector integration, mentioned 
above.  This elastic Green function will be derived in 
App.~\ref{sec:EGF}.  In terms of the elastic Green function and 
the force density, one finds for the increase $F[{\bf f}]-F[{\bf 0}]$ 
in free energy due to the applied force-density field:
\begin{subequations}
\begin{eqnarray}
F[{\bf f}]-F[{\bf 0}]
&=&
-T\lim_{n\to 0}
\frac{\rpf_{1+n}[{\bf f}]-\rpf_{1+n}[{\bf 0}]}
     {n\,\rpf_{1}[{\bf 0}]}
=-T\lim_{n\to 0}
\frac{\partial}{\partial n}\,
\ln
\frac{\rpf_{1+n}[{\bf f}]}
     {\rpf_{1+n}[{\bf 0}]}
\\
&\approx&
-\frac{1}{2\stiff_{0}}
\int_{\cal V} d{\bf x}\,d{\bf x}^{\prime}\,
\sum_{d,d^{\prime}=1}^{D}
f_{d}({\bf x})\,
{\cal G}_{dd^{\prime}}({\bf x}-{\bf x}^{\prime})\,
f_{d^{\prime}}({\bf x}^{\prime}),
\label{eq:freeenergystress}
\end{eqnarray}
\end{subequations}
which indicates that $\stiff_{0}$ is the shear modulus; 
see Refs.~\cite{LandauLifshitz,KleinertGFS}. 
\end{widetext}

\section{Effect of Goldstone fluctuations on the 
         order parameter and its correlations}
\label{sec:OPreduction}
In this section we discuss, how Goldstone fluctuations affect the
expectation value and the correlations of the order parameter.  Our
discussion is based on harmonic elasticity theory as described by the
free energy of eq.(\ref{eq:elasticenergy}). We shall make frequent use
of the elastic Green function, which is defined in
eqs.(\ref{eq:greenKS}) and computed in App.~\ref{sec:EGF}. The effects
of Goldstone fluctuations are most striking in two dimensions, hence
the two-dimensional solid will be discussed separately in
sec.(\ref{sec:TwoDimensionRS}).

\subsection{Order parameter reduction due to Goldstone fluctuations}
Classically, the order parameter expectation value is given 
by Eq.~(\ref{eq:classicalstate}) or, equivalently, 
\begin{subequations}
\begin{eqnarray}
&&
\big\langle 
V\op({\bf x},k_{\rtr})
\big\rangle
\equiv
\big\langle 
\int\dbar{\bf k}_{\rm tot}\,
e^{-i{\bf k}_{\rm tot}\cdot{\bf x}}\,
V\op(k)
\big\rangle
\\
&&\qquad\qquad=
\sum_{{\bf k}_{\rm tot}}\,
e^{-i{\bf k}_{\rm tot}\cdot{\bf x}}\,
\big\langle 
\op(k)
\big\rangle
\approx
\cw(k_{\rtr}).
\label{eq:nophasefluctuationfactor}
\end{eqnarray}
\end{subequations}
The effect of Goldstone fluctuations on the order parameter
expectation value is estimated from the Gaussian 
theory~(\ref{eq:elasticenergy}), via which we compute the 
mean value of the distorted classical state:
\begin{subequations}
\begin{eqnarray}
\big\langle 
V\op({\bf x},k_{\rtr})
\big\rangle
&=&
\big\langle 
\int\dbar{\bf k}_{\rm tot}\,
e^{-i{\bf k}_{\rm tot}\cdot{\bf x}}\, 
V\op(k)
\big\rangle
\\
&=&
\sum_{{\bf k}_{\rm tot}}\,
e^{-i{\bf k}_{\rm tot}\cdot{\bf x}}\,
\big\langle
\op(k)
\big\rangle
\label{eq:phasefluctuationop}
\\
&\approx&
\big\langle
e^{ik_{\rtr}\cdot u_{\rtr}({\bf x})}
\big\rangle\,
\cw(k_{\rtr}).
\label{eq:phasefluctuationfactor}
\end{eqnarray}
\end{subequations}
This is readily evaluated via the Gaussian property of $u$,
and hence we find
\begin{subequations}
\label{eq:gaussianreduction}
\begin{eqnarray}
\big\langle V\op({\bf x},k_{\rtr})\big\rangle
&\approx&
\widetilde{\cw}(k_{\rtr}),
\\
\widetilde{\cw}(k_{\rtr})
&\equiv&
\exp\left(-T\,\GatZ_{\! D}\,k_{\rtr}^{2}/2\stiff_{0}\right)\,
\cw(k_{\rtr}),
\label{eq:rendistdef}
\end{eqnarray}
\end{subequations}
for the fluctuation-renormalized form of $\cw(k_{\rtr})$.  Here
$D\,\GatZ_{\! D}\equiv{\cal G}_{dd}({\bf x})\vert_{{\bf x}={\bf 0}}$
and summation over repeated cartesian indices is implied.  Recalling
the classical structure for $\cw$, Eq.~(\ref{eq:mfstate}), we see that
the effect of the fluctuations is to induce, in say
Eq.~(\ref{eq:classicalstate}) or (\ref{eq:nophasefluctuationfactor}),
the replacement
\begin{subequations}
\label{eq:bundleshift}
\begin{eqnarray}
\cw(k_{\rtr})&=&
Q\int_{0}^{\infty}\!\!\! d\xi^{2}\,
\dsl\left(\xi^{2}\right)\,
e^{-\xi^{2}k_{\rtr}^{2}/2}
\longrightarrow
\widetilde{\cw}(k_{\rtr}), 
\\
\widetilde{\cw}(k_{\rtr})
&\equiv&
\,e^{-T\,\GatZ_{\! D}k_{\rtr}^{2}/2\stiff_{0}}\,
Q\int_{0}^{\infty}\!\!\! d\xi^{2}\,\dsl\left(\xi^{2}\right)\,
e^{-\xi^{2}k_{\rtr}^{2}/2}
\nonumber
\\
&=&
Q\,\int_{T\,\GatZ_{\! D}/\stiff_{0}}^{\infty}
\!\!\!\!\!\! d\xi^{2}\,
\dsl\left(\xi^{2}-(T\,\GatZ_{\! D}/\stiff_{0})\right)\,
e^{-\xi^{2}k_{\rtr}^{2}/2}
\nonumber
\\
&=&
Q\,\int_{0}^{\infty}d\xi^{2}\,
\wt{\dsl}\left(\xi^{2}\right)\,
e^{-\xi^{2}k_{\rtr}^{2}/2},
\label{eq:shiftindistribution}
\end{eqnarray}
\end{subequations}
i.e., a rigid shift of the distribution $\dsl$ to longer (squared) 
localization lengths, as encoded in the new distribution $\wt{\dsl}$.  
This is to be expected: the locally fluctuating localized objects 
are also subject to collective fluctuations---phonons.

This shift of $\dsl$ is determined by the value of $\GatZ_{\! D}$ 
which, as shown in App.~\ref{sec:EGF}, has the following 
leading-order dependence on $D$: 
\def\DPS{\displaystyle}
\begin{equation}
\GatZ_{\! D}\approx
\begin{cases}
\frac{\DPS\Sigma_{D}}{\DPS(2\pi)^{2}}
\frac{\DPS D-1}{\DPS D(D-2)}\,\,
\frac{\DPS 1}{\DPS\coshort^{D-2}},
&{\rm for}\quad D>2; 
\\
\noalign{\medskip}
\frac{1}{4\pi}\ln\big(\colong/\coshort\big),
&{\rm for}\quad D=2.
\end{cases}
\label{eq:casesofGzero}
\end{equation}
where $\Sigma_{D}$ is the area of the $(D-1)$-dimensional
surface of a $D$-dimensional sphere of radius unity, viz, 
\begin{equation}
\Sigma_{D}=2\pi^{D/2}/\Gamma(D/2),  
\label{eq:SurfaceD}
\end{equation} 
in which $\Gamma(D)$, 
[not to be confused with $\GatZ_{\! D}$ 
of Eq.~(\ref{eq:casesofGzero})] 
is the conventional Gamma function.
We see that in dimension $D$ greater than two the shift is 
{\it finite\/}.  However, at and below $D=2$ the shift 
{\it diverges} 
with the long-distance cut-off $\colong$, viz., the linear size 
of the system, doing so logarithmically in two dimensions.  
Not surprisingly, fluctuations destroy particle localization in 
two-dimensional amorphous solids and thus restore the symmetry 
broken at the classical level (see below).

We now consider the scaling behavior of the fluctuation-induced 
shift of the distribution from $\dsl$ to $\wt\dsl$.  
We see from Eqs.~(\ref{eq:bundleshift}) tha this shift is 
parametrized by the length $\llshift$, which is defined via 
\begin{equation}
\llshift^{2}\equiv\GatZ_{\! D}\,T/\stiff_{0}\,.
\label{eq:FluctLength}
\end{equation}
How does this length compare with the typical localization length $\lltyp$, 
as defined, say, via the most probable value of $\xi$ predicted by the 
classical theory?  From Eq.~(\ref{eq:distribution}) we see that  
\begin{equation}
\lltyp^{2}\sim\lsz^{2}/\tau, 
\label{eq:Typical}
\end{equation}
and thus we have that 
\begin{equation}
\llshift^{2}/\lltyp^{2}
\sim
\GatZ_{\! D}\,T\,\tau/\stiff_{0}\,\lsz^{2}\,. 
\label{eq:lengthquotient}
\end{equation}
Now, from Eq.~(\ref{eq:shearscale}) we have that  
\begin{equation}
\stiff_{0}/T\sim\den\,\tau^{3}, 
\end{equation}
and we use this to eliminate $\stiff_{0}/T$ in 
Eq.~(\ref{eq:lengthquotient}).  Furthermore, from 
Eq.~(\ref{eq:casesofGzero}) we have that, for $D>2$,  
\begin{equation}
\GatZ_{\! D}\sim \coshort^{2-D}\sim\lltyp^{2-D}, 
\end{equation}
provided we take the short-distance cut-off to be of order $\lltyp$, 
as discussed in Sec.~\ref{sec:ShearModulus}.
We use this to eliminate $\GatZ_{\! D}$ in favor of $\lltyp$.  
Finally, by using Eq.~(\ref{eq:Typical}) to eliminate $\tau$ 
in favor of $\lltyp/\lsz$ we obtain, for $D>2$,  
\begin{equation}
{\llshift^{2}}\,/\,\lltyp^{2}\sim 
\big(c\,\lsz^{D}\big)^{-1}\,
\big(\lltyp/\lsz\big)^{6-D}.
\label{eq:HowToScale}
\end{equation}
At $D=6$ we find {\it one-parameter scaling\/} in the sense that 
$\llshift\sim\lltyp$, up to the factor $c\,\lsz^{D}$, which measures 
the number of crosslinked entities within a region of order the size 
of a single one.   This is to be expected, because we have used 
classical exponents for the divergence of $\lltyp$ and vanishing of 
$\stiff_{0}$ with $\tau$, and six is the upper critical dimension for 
the transition to the amorphous solid state.  In fact, if we were to 
replace these classical exponent by their anomalous values we would 
expect to recover one-parameter scaling at arbitrary $D$.
\begin{widetext}
\subsection{Two-field order parameter correlations}
\label{sec:EVtwoFC}
As with their effect on the order parameter itself, the effect 
of Goldstone fluctuations on the two-field correlator can be 
determined from the Gaussian theory~(\ref{eq:elasticenergy}), 
which gives 
\begin{equation}
\big\langle
V\op({\bf x},k_{\rtr})\,
V\op({\bf x}^{\prime},k_{\rtr}^{\prime})^{\ast}
\big\rangle
\approx
\big\langle 
e^{ ik_{\rtr}\cdot u_{\rtr}({\bf x})}\,
e^{-ik_{\rtr}^{\prime}\cdot u_{\rtr}({\bf x}^{\prime})}
\big\rangle\,
\cw(k_{\rtr})\,\cw(k_{\rtr}^{\prime}).
\label{eq:correlatorfactor}
\end{equation}
The required correlator 
$\big\langle
\exp  ik_{\rtr}         \cdot u_{\rtr}({\bf x}         )\,
\exp -ik_{\rtr}^{\prime}\cdot u_{\rtr}({\bf x}^{\prime})
\big\rangle$
is also readily evaluated via the gaussian property of $u$, and 
hence we have 
\begin{equation}
\Big\langle
V\op({\bf x},k_{\rtr})\,
V\op({\bf x}^{\prime},k_{\rtr}^{\prime})^{\ast}
\Big\rangle
\approx
\exp\left(
{-\frac{T\,\GatZ_{\! D}}{2\stiff_{0}}
\vert{k_{\rtr}-k_{\rtr}^{\prime}}\vert^{2}}\right)
\exp\left({-\frac{T}{\stiff_{0}}\,
\big({\cal G}_{dd^{\prime}}({\bf 0})-
      {\cal G}_{dd^{\prime}}({\bf x}-{\bf x}^{\prime})\big)
k_{\rtr d}^{\phantom\prime}\cdot k_{\rtr d^{\prime}}^{\prime}}\right)\,
\cw(k_{\rtr})\,\cw(k_{\rtr}^{\prime}),
\label{eq:correlatorvalue}
\end{equation}
where the scalar products 
$k_{\rtr d}^{\phantom\prime}\cdot k_{\rtr d^{\prime}}^{\prime}$ and 
$\vert{k_{\rtr}-k_{\rtr}^{\prime}}\vert^{2}$
are respectively taken over $n$ and $nD$ components. 

\end{widetext}
Recall, from Eq.~(\ref{eq:casesofGzero}), that $\GatZ_D\vert_{D=2}$ 
diverges with the long-distance cut-off.  This, together with the 
positive-semi-definiteness of the quadratic form 
$\big({\cal G}_{dd^{\prime}}({\bf 0})-
      {\cal G}_{dd^{\prime}}({\bf x}-{\bf x}^{\prime})\big)
      k_{\rtr d}\cdot k_{\rtr d^{\prime}}^{\prime}$ in 
Eq.~(\ref{eq:correlatorvalue}), makes it evident that the 
correlator 
$\big\langle
V\op({\bf x},k_{\rtr})\,
V\op({\bf x}^{\prime},k_{\rtr}^{\prime})^{\ast}
\big\rangle$ 
given in Eq.~(\ref{eq:correlatorvalue}) vanishes at $D=2$ 
(and below) unless $k=k^{\prime}$.  This vanishing is a 
second facet of the fluctuation-induced restoration of symmetry 
discussed for the case of the order parameter, following 
Eq.~(\ref{eq:casesofGzero}).  [The correlator 
$\big\langle V\op(k)\,V\op(k^{\prime})^{\ast}\big\rangle$ 
vanishes in 
{\it any\/} dimension unless $k_{\rlo}=k_{\rlo}^{\prime}$,
owing to the preserved symmetry of common translations of the 
replicas, which encodes the homogeneity of the randomness in 
the amorphous solid state, i.e., its macroscopic translational 
invariance.]\thinspace\  If $k=k^{\prime}$ then, regardless of 
dimension, the correlator decays with increasing separation 
${\bf x}-{\bf x}^{\prime}$, as shown by Eq.~(\ref{eq:correlatorvalue}).

For $D>2$ it is convenient to analyze the two-field correlator
normalized by its disconnected part, i.e.,
\begin{equation}
\frac{\DPS\big\langle\op({\bf x},k_{\rtr})\,
\op({\bf x}^{\prime},k_{\rtr}^{\prime})^{\ast}\big\rangle}
{\DPS\big\langle\op({\bf x},k_{\rtr})\big\rangle\,\big\langle
\op({\bf x}^{\prime},k_{\rtr}^{\prime})^{\ast}\big\rangle}.
\end{equation}
Equations~(\ref{eq:correlatorvalue}) and (\ref{eq:gaussianreduction})
show this quotient to be given by
\begin{equation}
\exp\left(\frac{T}{\stiff_{0}}\,\,
  {\cal G}_{dd^{\prime}}({\bf r})\,
k_{\rtr d}\cdot
k_{\rtr d^{\prime}}^{\prime}\right), 
\label{eq:correlatorquotient}
\end{equation}
where the separation ${\bf r}\equiv{\bf x}-{\bf x}^{\prime}$.

We illustrate this behavior by considering the case of $D=3$ and 
the regime $\coshort\ll\vert{{\bf r}}\vert\ll\colong$, for which, 
as shown in App.~\ref{sec:EGF} (see also, e.g., 
Ref.~\cite{KleinertGFS}), we may use 
\begin{equation}
{\cal G}_{dd^{\prime}}^{(3)}({\bf r})
\approx 
\frac{1}{8\pi\,\vert{\bf r}\vert}
\left(
 \delta_{dd^{\prime}}
+\hat{r}_{d}\,\hat{r}_{d^{\prime}}
\right),
\label{eq:threeDGreen}
\end{equation}
where the unit vector 
$\hat{\bf r}\equiv{\bf r}/\vert{\bf r}\vert$, 
and by choosing
$k=k^{\prime}=\{{\bf 0},{\bf q},-{\bf q},{\bf 0},\ldots,{\bf 0}\}$.
Then in this regime the normalized two-field correlator is given by
\begin{equation}
\exp
\left(
\frac{T}{4\pi\stiff_{0}}
\frac{\vert{\bf q}\vert^{2}}
     {\vert{\bf r}\vert^{\phantom{2}}}\,
     [1+\cos^{2}\!\varphi]
\right),
\label{eq:correlator3Dexamples}
\end{equation}
which depends strongly on the angle $\varphi$
between the vectors ${\bf q}$ and ${\bf r}$. 

We may also examine the normalized two-field correlator in the regime 
$\vert{\bf r}\vert\ll\coshort$.   Making use of  
${\cal G}_{dd^{\prime}}^{(3)}({\bf r})$ in this regime, as given in 
Eq.~(\ref{eq:GshortThree}) of App.~\ref{sec:EGF}, we find that the 
normalized two-field correlator is given by 
\begin{equation}
\exp
\Big(
\frac{4\,T}{3\pi\stiff_{0}}
\frac{\vert{\bf q}\vert^{2}}{\coshort}
\Big)\,
\exp
\Big(
\frac{2\,T}{45\pi\stiff_{0}}
\frac{\vert{\bf q}\vert^{2}}{\coshort}
\frac{\vert{\bf r}\vert^{2}}{\cbshort^{2}}\,
[-2+\cos^{2}\!\varphi]
\Big), 
\label{eq:correlator3Dshort}
\end{equation}
where 
$\cbshort\equiv\coshort/2\pi$ (and 
$\cblong \equiv\colong /2\pi$). 
Of course, in this regime the result for the correlator is incomplete, 
as there will also be contributions from the non-Goldstone excitations. 

\subsection{Intermezzo on length-scales}
\label{sec:Lengthscales}
We pause to catalog the various length-scales featuring in the 
present Paper, and to indicate where they first appear.  

The shortest length is $\lpc$, the persistence length of a macromolecule, 
which appears alongside $L$, the arclength of a macromolecule, shortly 
after Eq.~(\ref{eq:Landaueffective}).  
Together, they yield $\lsz$, the linear size of the objects being linked, 
which may be substantially larger than $\lpc$.  It is only through $\lsz$ 
that $\lpc$ and $L$ feature in the free energy, 
Eq.~(\ref{eq:Landaueffective}).
The density $\den$ of entities being constrained also first features in 
Eq.~(\ref{eq:Landaueffective}), and sets a length-scale, $\den^{-1/D}$, 
in Eq.~(\ref{eq:HowToScale}).
Comparable to or longer than $\lsz$ is the length $\xi$, the 
(distributed) localization length, first featuring in 
Eqs.~(\ref{eq:RMSflucts}) and (\ref{eq:defdistrib}).  
The typical value of $\xi$ is denoted $\lltyp$, first mentioned 
around Eq.~(\ref{eq:Typical}). 
The amount by which fluctuations shift the distribution of $\xi$ is 
encoded in the fluctuation length $\llshift$, first mentioned around 
Eq.~(\ref{eq:FluctLength}). 
The short- and long-distance cut-offs for the Goldstone-type fluctuations 
are respectively denoted $\coshort$ and $\colong$, and are first mentioned 
in Sec.~\ref{sec:ShearModulus}.  
The linear size of the sample is denoted $\linsize$; see the beginning of 
Sec.~\ref{sec:TwoDimensionRS}.  

\section{Amorphous solids in two dimensions}
\label{sec:TwoDimensionRS}
We now focus on amorphous solids in {\it two\/} dimensions. 
By taking the long-distance cut-off $\colong$ to be the linear 
size of the sample $\linsize$, so that its area is 
$\linsize^{2}$, we see from Eqs.~(\ref{eq:gaussianreduction}) 
and (\ref{eq:casesofGzero}) that the expectation value of the 
order parameter does indeed 
{\it vanish logarithmically\/} with the sample size, doing 
with an exponent that depends, inter alia, on $k_{\rtr}$:
\begin{eqnarray}
\big\langle 
\linsize^{2}\,
\op({\bf x},k_{\rtr})\big\rangle
&\approx&
\cw(k_{\rtr})\,
\exp\left(-
\smhalf
\GMZeta(k_{\rtr}^{2})
\ln\big(\linsize/\coshort\big)\right) 
\nonumber\\
&&\qquad=
\cw(k_{\rtr})\,
\big(\linsize/\coshort\big)^{-\GMZeta(k_{\rtr}^{2})/2}, 
\label{eq:logsuppression}
\end{eqnarray}%
where the exponent $\GMZeta(\kappa^{2})$ varies continuously 
with wave-number and is defined via
\begin{equation}
\GMZeta(\kappa^{2})\equiv\frac{T\kappa^{2}}{4\pi\stiff_{0}}.
\end{equation}
To arrive at this result we have made use of the two-dimensional 
real-space elastic Green function evaluated at the origin, computed 
in App.~\ref{sec:EGF} and stated in Eq.~(\ref{eq:casesofGzero}). 
It confirms the expectation, mentioned above, that in two-dimensions 
fluctuations destroy particle localization and restore the broken symmetry.  
Note, however that, due to the purely entropic nature of the 
elasticity the exponent does not depend on temperature 
[cf.~Eq.~(\ref{eq:shearscale})]. 

Similarly, from Eq.~(\ref{eq:correlatorvalue}) we see that the
$k_{\rtr}$-diagonal two-field correlator is given by
\begin{eqnarray}
&&\big\langle
\linsize^{2}\,\op({\bf x},k_{\rtr})\,
\linsize^{2}\,\op({\bf x}^{\prime},k_{\rtr})^{\ast}
\big\rangle\approx
\label{eq:diagcorrelator}
\\ 
&&
\cw(k_{\rtr})^{2}\,
\exp
\left(
{-\frac{T}{\stiff_{0}}
\big({\cal G}_{dd^{\prime}}({\bf 0})-
     {\cal G}_{dd^{\prime}}({\bf r})\big)
      k_{\rtr d}\cdot k_{\rtr d^{\prime}}} 
\right), 
\nonumber
\end{eqnarray}
where, as before, the separation is denoted by 
${\bf r}\equiv{\bf x}-{\bf x}^{\prime}$ 
and the unit vector by 
$\hat{\bf r}\equiv{\bf r}/\vert{\bf r}\vert$. 
In the regime 
$\coshort\ll\vert{\bf r}\vert\ll\linsize$, 
the Green function is given by 
[see Eq.~(\ref{eq:EGFtwoDint}) in App.~\ref{sec:EGF}]
\begin{equation}
{\cal G}_{dd^{\prime}}({\bf 0})-
{\cal G}_{dd^{\prime}}({\bf r})=
\frac{1}{4\pi}
\left\{\delta_{dd^{\prime}}
  \ln\big(r/\,\cRshort\big)
-\hat{r}_d\,\hat{r}_{d^{\prime}}
\right\}, 
\end{equation}
where, for convenience, we have introduced the numerically 
rescaled short-distance cut-off $\cRshort$, defined via 
\begin{equation}
\cRshort\equiv\frac{\coshort}{\pi e^{\gamma+\frac{1}{2}}}
\end{equation}
In consequence, the correlator decays algebraically with the 
{\it magnitude\/} $r$ of the distance, and with an amplitude 
that depends on the {\it orientation\/} $\hat{\bf r}$ of the 
distance (relative to $k_{\rtr}$).  Ignoring, for the moment 
the dependence on $\hat{\bf r}$, we have the leading-order 
result:
\begin{equation}
\big\langle
\linsize^{2}\,\op({\bf x},k_{\rtr})\,
\linsize^{2}\,\op({\bf x}^{\prime},k_{\rtr})^{\ast}
\big\rangle\approx
\cw(k_{\rtr})^{2}\,
\left(\cRshort/r\right)^{\GMZeta(k_{\rtr}^{2})}.
\label{EQ:RangeOnly}
\end{equation}
As mentioned above, the $k_{\rtr}$-off-diagonal two-field 
correlator---like the order parameter---vanishes in the 
thermodynamic limit, reflecting the fluctuation-induced 
restoration of symmetry.  

This scenario is similar to the theory of two-dimensional regular
solids, taking into account harmonic phonons only, as discussed in
Ref.~\cite{Jancovici}.  The main difference is the absence of Bragg
peaks at reciprocal lattice vectors. Instead, there is a divergence 
in the scattering function~\cite{Jancovici} at zero wave-number, 
as can be seen from the Fourier transform of the 
correlator~(\ref{EQ:RangeOnly}) with respect to 
${\bf r}$ ($\equiv{\bf x}-{\bf x}^{\prime}$):
\begin{eqnarray}
&&
S({\bf q},k_{\rtr})=
\int d{\bf r}\,
e^{i{\bf r}\cdot{\bf q}}\,
\big\langle
\linsize^{2}\,\op({\bf x},k_{\rtr})\,
\linsize^{2}\,\op({\bf x}^{\prime},k_{\rtr})^{\ast}
\big\rangle
\nonumber\\
&&
=\cw(k_{\rtr})^{2}\,
\coshort^{2}
\left\{
\left(\frac{\colong}{\coshort}\right)^{2-\GMZeta}
\frac{{\rm J}_{1}(q\colong)}{q\colong}
+B\big(\GMZeta\big)\,
\big(q\coshort\big)^{\GMZeta-2}
\right\},
\nonumber\\
&&
\qquad\qquad
B(x)\equiv 2^{-x}\,
\frac{x \Gamma(1-x/2)}{\phantom{x}\Gamma(1+x/2)}.
\end{eqnarray}
In these formul\ae, ${\rm J}_1$ denotes a Bessel function and 
we have abbreviated $\GMZeta(k_{\rtr}^{2})$ as $\GMZeta$.  
One needs to keep the two exhibited terms because there is an 
exchange of dominance as $\GMZeta$ passes through the value 1/2, 
i.e., at $k_{\rtr}^{2}=2\pi\stiff_{0}/T$.  
Specifically, for $(\colong/\coshort)\to\infty$ at fixed $q\coshort$ 
the dimensionless factor $\{\cdots\}$ behaves as: 
\begin{subequations}
\begin{eqnarray}
&&(q\coshort)^{-3/2}\,
(\colong/\coshort)^{\smhalf-\GMZeta},
\quad
{\rm for\/}\quad\GMZeta<1/2;
\\
&&(q\coshort)^{\GMZeta-2},\,
\phantom{(\colong/\coshort)^{\smhalf-\GMZeta}\,\,}
\quad
{\rm for\/}\quad\GMZeta>1/2.
\end{eqnarray}
\end{subequations}

Restoring the dependence on orientation $\hat{\bf r}$, we arrive at 
the full, anisotropic form for the decay of the two-field correlator: 
\begin{equation}
\cw(k_{\rtr})^{2}
  \left({\cRshort}/r\right)^{\GMZeta(k_{\rtr}^{2})}
\exp
\left(
\GMZeta(k_{\rtr}^{2})\,
{\hat k}_{\rtr d_{1}}\cdot
{\hat k}_{\rtr d_{2}}\,
{\hat{r}}_{d_{1}}
{\hat{r}}_{d_{2}}
\right).
\label{eq:algebraicdecaygeneral}
\end{equation}
Here, the symbol 
${\hat k}_{\rtr d}$ indicates the unit vector
${k}_{\rtr d}/\vert{{k}_{\rtr}}\vert$.
For the illustrative case of
$k=({\bf 0},{\bf q},-{\bf q},{\bf 0},\ldots)$
this correlator reduces to
\begin{equation}
\cw\Big(\sqrt{2}\,\vert{\bf q}\vert\Big)^{2}
\left({\cRshort}/r\right)^{2\GMZeta(q^{2})}
\exp\left(2\GMZeta(q^{2})\,\cos^{2}\!\varphi\right), 
\label{eq:algebraicdecayspecial}
\end{equation}
where, once again, $\varphi$ is the angle 
between ${\bf q}$ and ${\bf r}$. 

The framework adopted in the present Paper allows the identification of the 
{\it quasi-localized fraction\/} $\qlf$, i.e., the fraction of particles 
that, whilst not truely localized, have R.M.S.~displacements that diverge 
only logarithmically, as the system size $\linsize$ goes to infinity.  
This fraction should be contrasted with the delocalised fraction, i.e., 
the fraction that have R.M.S.~displacements that diverge linearly with 
$\linsize$.  One way to extract the quasi-localized fraction is via the 
order parameter~(\ref{eq:logsuppression})---specifically the 
amplitude of its power-law decay with $\linsize$.  A simpler route is 
via the two-field correlator~(\ref{eq:algebraicdecaygeneral}):
\begin{eqnarray}
\qlf^{2}\!\!&=&\!\!
\lim_{k_{\rtr}\to 0}
\lim_{r\to \infty}
\left({\cRshort}/r\right)^{-\GMZeta(k_{\rtr}^{2})}
\!\big\langle
\linsize^{2}\,\op({\bf x},k_{\rtr})\,
\linsize^{2}\,\op({\bf x}^{\prime},k_{\rtr})^{\ast}
\big\rangle
\nonumber\\
&=&\cw(0)^{2}.
\label{eq:quasolocal}
\end{eqnarray}

The option of analyzing the quasi-localized fraction would not be 
available to percolation-based approaches to vulcanized matter, as such 
approaches do not account for the thermal motion of the constituents 
(e.g.~macromolecules), but only the architecture of the structures 
they constitute.  In fact, for the same reason, the entire circle of 
ideas described in the present Paper lie beyond the reach of 
percolation-based approaches.  This shows up especially vividly in the 
context of the vulcanization transition in low (and especially two) 
dimensions.  Whereas percolation theory would indicate a lower critical 
dimension of unity, with nothing fundamentally new happening as one 
reduces the dimensionality through two, the present approach correctly 
finds a lower critical dimension at two dimensions for the amorphous 
solidification transition.  
This is because, in addition to incorporating the physics of 
percolation~\cite{Peng+Goldbart2000,Peng+Goldbart+McKane2001}, this 
approach contains the logically-independent physics of localization and 
the attendant issue of the spontaneous breaking of translational 
symmetry.  Thus, there are qualitative, and not merely quantitative, 
distinctions between the vulcanization transition and resulting 
quasi-amorphous solid state in two- and in higher-dimensional settings, 
owing to the strong role played by Goldstone-type fluctuations in 
reduced dimensions.

The results discussed in the present section echo those found for 
a variety of two-dimensional statistical-mechanical systems (for a 
survey see, e.g., Ref.~\cite{Nelson2001}), in the following sense. 
A sufficient density of random constraints triggers a phase 
transition from a liquid state to a quasi-amorphous solid state.  
In the liquid state there are no clusters of constituents that 
span the system (i.e.~the system does not percolate), there are 
no quasi-localized particles, order-parameter correlations decay 
exponentially with distance, and the static shear modulus is zero.  
In the solid state a cluster does span the system (i.e.~percolates), 
there are quasi-localized particles, the static shear modulus is 
non-zero, and order-parameter correlations decay alebraically, with 
a continuously-varying exponent $\GMZeta$ that depends, inter alia, 
on the scale set by the {\it probe wave-vector\/} $k_{\rtr}$.  
However, the true long-range order found in higher dimensions, 
fails to set in, albeit only just, due to thermal fluctuations. 

The situation is reminiscent of that found in the setting of the 
melting of two-dimensional crystals, triggered not by the density 
of constraints, but rather by thermal fluctuations.  In this setting, 
at high temperatures one has a non-rigid phase with exponentially 
decaying positional correlations associated with unbound dislocations.  
This is the analog of the non-percolating state of quasi-amorphous 
solids.  At lower temperatures one has a rigid phase, associated with 
bound dislocations and algebraically decaying positional correlations, 
but no true long-range positional order.  This is the analog of the 
percolating regime of amorphous solids.  The permanence of the 
architecture of the amorhphous solid forming systems discussed in 
the present Paper prohibits the destruction by thermal fluctuations 
of a percolating raft of constituents.  However, percolation does 
not enforce true localization.

In two-dimensional crystallization one also has the opportunity 
for correlations in the {\it orientations\/} of the bonds connecting 
neighboring particles.  Thus, at low temperaures one has, in 
addition to quasi-long range positional correlations, true long-range 
orientational correlations.  This brings the opportunity for an 
intermediate regime of temperatures, in which dislocations are 
unbound but disclinations, which would disrupt the orientational 
order, remain bound: the hexatic phase of two-dimensional crystals. 

This opportunity does not seem to be present in the model of 
vlucanized matter discussed here, which focuses exclusively on 
positional order and does not support topological excitations.  
Richer settings, such as those involving macromolecules with liquid 
crystalline degrees of freedom, seem likely to raise interesting 
opportunities for order in low-dimensional random systems.

\begin{widetext}
\section{Physical content of correlators}
\label{sec:OPcorrelators}
\subsection{Identifying the statistical information 
in the two-field correlator}
\label{sec:identify}
What is the meaning of the correlators in the amorphous solid state 
of the vulcanization field theory?  
Let us begin with the two-field correlator
$\langle\op(k_{1})\,\op(k_{2})^{\ast}\rangle$.  
Specializing to the case in which the zero-replica entries
${\bf k}_{1}^{0}$ and ${\bf k}_{2}^{0}$ 
are zero, the interpretation of this replica quantity is given by
\begin{equation}
\Big\langle\op(k_{1})\,\op(k_{2})^{\ast}\Big\rangle
\Big\vert_{{\bf k}_{1}^{0}={\bf k}_{2}^{0}={\bf 0}}=
\Big[
\frac{1}{J^{2}}
\sum_{j_{1},j_{2}=1}^{J}
\prod_{\alpha=1}^{n}
\left\langle
e^{i{\bf k}_{1}^{\alpha}\cdot{\bf R}_{j_{1}}
  -i{\bf k}_{2}^{\alpha}\cdot{\bf R}_{j_{2}}}
\right\rangle
\Big], 
\label{eq:repint}
\end{equation}
which expresses the connection with the (semi-microscopic) correlators 
of all pairs of particles that constitute the system.  Consider a 
simple illustrative example, in which the positions of pairs of 
localized particles fluctuate about their mean positions according to 
a general gaussian correlated distribution.  For such particles we 
would then have
\begin{eqnarray}
\left\langle
e^{i{\bf k}_{1}\cdot{\bf R}_{j_{1}}
  -i{\bf k}_{2}\cdot{\bf R}_{j_{2}}}
\right\rangle&=&
e^{ i k_{1d}r_{j_{1}d}
   -i k_{2d}r_{j_{2}d}}
e^{-\frac{1}{2}
    \Delta r_{j_{1}j_{1}d_{1}d_{2}}
     k_{1d_{1}}k_{1d_{2}}}
e^{-\frac{1}{2}
    \Delta r_{j_{2}j_{2}d_{1}d_{2}}
     k_{2d_{1}}k_{2d_{2}}}
e^{ \Delta r_{j_{1}j_{2}d_{1}d_{2}}
     k_{1d_{1}}k_{2d_{2}}}
\end{eqnarray}
where the mean, variances ($j_{1}=j_{2}$) and co-variances
($j_{1}\ne j_{2}$) of the positions are given by
\begin{subequations}
\begin{eqnarray}
{\bf r}_{j}&=&\langle{\bf R}_{j}\rangle,
\\
\Delta{r}_{j_{1}j_{2}d_{1}d_{2}}&=&
\langle
\left({R_{j_{1}d_{1}}-\langle{R_{j_{1}d_{1}}}\rangle}\right)
\left({R_{j_{2}d_{2}}-\langle{R_{j_{2}d_{2}}}\rangle}\right)
\rangle,
\end{eqnarray}
\end{subequations}
and summations over repeated cartesian indices 
$d_{1}$ and $d_{2}$ are implied. 

Putting such contributions together from all pairs of 
{\it localized\/}, particles we have
\begin{eqnarray}
&& 
\Big\langle
\op(k_{1})\,\op(k_{2})^{\ast}
\Big\rangle
\Big\vert_{{\bf k}_{1}^{0}={\bf k}_{2}^{0}={\bf 0}}
=Q^{2}
\int d\nh{\bf r}_{1}\,d\nh{\bf r}_{2}\,
     d\wh{\Delta r}_{11}\,d\wh{\Delta r}_{22}\,d\wh{\Delta r}_{12}\,
\cpd(\nh{\bf r}_{1},\nh{\bf r}_{2},
\wh{\Delta r}_{11},\wh{\Delta r}_{22},\wh{\Delta r}_{12})\,
\nonumber\\ &&\qquad\times
e^{i\nh{r}_{1d}\sum_{\alpha=1}^{n}k_{1d}^{\alpha}
  -i\nh{r}_{2d}\sum_{\alpha=1}^{n}k_{1d}^{\alpha}}
e^{-\frac{1}{2}\wh{\Delta r}_{11d_{1}d_{2}}
    \sum_{\alpha=1}^{n}k_{1d_{1}}^{\alpha}k_{1d_{2}}^{\alpha}}
e^{-\frac{1}{2}\wh{\Delta r}_{22d_{1}d_{2}}
    \sum_{\alpha=1}^{n}k_{2d_{1}}^{\alpha}k_{2d_{2}}^{\alpha}}
e^{            \wh{\Delta r}_{12d_{1}d_{2}}
    \sum_{\alpha=1}^{n}k_{1d_{1}}^{\alpha}k_{2d_{2}}^{\alpha}},
\label{eq:CorInProb}
\\
&&
\!\!\!\!
\cpd(\nh{\bf r}_{1},\nh{\bf r}_{2},
\wh{\Delta r}_{11},
\wh{\Delta r}_{22},
\wh{\Delta r}_{12})
\!\equiv\!\!\left[
\frac{1}{(QJ)^{2}}
\!\!\!\!\sum_{j_{1},j_{2}\,{\rm loc.}}\!\!\!\!\!
\delta(\nh{\bf r}_{1}-{\bf r}_{j_{1}})
\delta(\nh{\bf r}_{2}-{\bf r}_{j_{2}})
\delta(\wh{\Delta r}_{{1}{1}}-\Delta r_{j_{1}j_{1}})
\delta(\wh{\Delta r}_{{2}{2}}-\Delta r_{j_{2}j_{2}})
\delta(\wh{\Delta r}_{{1}{2}}-\Delta r_{j_{1}j_{2}})
\right]\!\!,
\nonumber
\end{eqnarray}
where we have introduced $\cpd$, the disorder-averaged joint 
distribution, over the pairs of localized particles, of the means, 
variances and co-variances of the particle positions.  The 
macroscopic translational invariance 
of the amorphous solid state ensures that $\cpd$ depends on 
$\nh{\bf r}_{1}$ and 
$\nh{\bf r}_{2}$ only through their difference, 
and thus we replace $\cpd$ by 
$V^{-1}\cpd(\nh{\bf r}_{1}-\nh{\bf r}_{2},
            \wh{\Delta r}_{11},\wh{\Delta r}_{22},\wh{\Delta r}_{12})$.
By appealing to permutation symmetry, including the 
zeroth replica, we can reinstate the dependence on the zeroth-replica 
wave vectors, and hence arrive at a hypothesized form for the two-point 
correlator:
\begin{eqnarray}
&&\!\!\!\!\!
\big\langle
\op(k_{1})\,\op(k_{2})^{\ast}
\big\rangle
=Q^{2}
\int d\nh{\bf r}_{1}\,d\nh{\bf r}_{2}\,
     d\wh{\Delta r}_{11}\,d\wh{\Delta r}_{22}\,d\wh{\Delta r}_{12}\,
     V^{-1}\,
\cpd(\nh{\bf r}_{1}-\nh{\bf r}_{2},
\wh{\Delta r}_{11},\wh{\Delta r}_{22},\wh{\Delta r}_{12})\,
\nonumber\\ &&\!\!\!\!\!\quad\times
e^{i\nh{r}_{1d}\sum_{\alpha=0}^{n}k_{1d}^{\alpha}
  -i\nh{r}_{2d}\sum_{\alpha=0}^{n}k_{1d}^{\alpha}}
e^{-\frac{1}{2}\wh{\Delta r}_{11d_{1}d_{2}}
    \sum_{\alpha=0}^{n}k_{1d_{1}}^{\alpha}k_{1d_{2}}^{\alpha}}
e^{-\frac{1}{2}\wh{\Delta r}_{22d_{1}d_{2}}
    \sum_{\alpha=0}^{n}k_{2d_{1}}^{\alpha}k_{2d_{2}}^{\alpha}}
e^{            \wh{\Delta r}_{12d_{1}d_{2}}
    \sum_{\alpha=0}^{n}k_{1d_{1}}^{\alpha}k_{2d_{2}}^{\alpha}}.
\label{eq:CorrelMeaning}
\end{eqnarray}

What about contributions to the double sum over particles in 
Eq.~(\ref{eq:repint}) associated with one or two {\it unlocalized\/} 
particles?  Such contributions certainly exist, except in the limit 
of large crosslink densities (where all constituents are bound to 
the infinite cluster and are, therefore, localized); in the liquid 
state they would, of course, be the only contributions.  There, they 
would give rise to {\it diagonal\/} contributions, i.e., the correlator 
$\big\langle\op(k_{1})\,\op(k_{2})^{\ast}\big\rangle$
would vanish unless $k_{1}=k_{2}$, owing to the intact symmetry of 
independent translations of the replicas.  In the solid state, 
contributions associated with unlocalized particles are expected 
to gives rise to short-ranged correlations.  As such correlations 
are not the main focus of the present paper, we neglect contributions 
to the two-field correlator associated with unlocalized particles.

\subsection{Evaluating the statistical information 
in the two-field correlator}
\label{sec:ImplTFC}
In Sec.~\ref{sec:identify} we have connected the two-field 
correlator 
$\big\langle V\op(k)\,V\op(k^{\prime})^{\ast}\big\rangle$ 
to the distribution $\cpd$ that characterizes pairs of 
localized particles [see Eq.~(\ref{eq:CorrelMeaning})], 
and in Sec.~\ref{sec:EVtwoFC} 
have evaluated the Fourier transform of this correlator, 
$\big\langle 
V\op({\bf x},k_{\rtr})\,
V\op({\bf x}^{\prime},k_{\rtr}^{\prime})^{\ast}
\big\rangle$, 
within the Goldstone-type fluctuation approach [see 
Eq.~(\ref{eq:correlatorvalue})].  We now discuss the implications 
of resulting correlator for the properties of the distribution.  

To do this, it is convenient to exchange the correlator in 
Eq.~(\ref{eq:correlatorvalue}) for the Fourier transform: 
\begin{eqnarray}
\big\langle\op(k)\,\op(k^{\prime})^{\ast}\big\rangle
&=&
\int
\frac{d{\bf x}}{V}\,
\frac{d{\bf x}^{\prime}}{V}\,
e^{ i{\bf k}_{\rm tot}\cdot{\bf x}}\,
e^{-i{\bf k}_{\rm tot}^{\prime}\cdot{\bf x}^{\prime}}\,
\big\langle
V\op({\bf x},k_{\rtr})\,
V\op({\bf x}^{\prime},k_{\rtr}^{\prime})^{\ast}
\big\rangle
\nonumber\\
&\approx&
\widetilde{\cw}(k_{\rtr})\,
\widetilde{\cw}(k_{\rtr}^{\prime})
\int
\frac{d{\bf x}}{V}\,
\frac{d{\bf x}^{\prime}}{V}\,
e^{ i{\bf k}_{\rm tot}\cdot{\bf x}}\,
e^{-i{\bf k}_{\rm tot}^{\prime}\cdot{\bf x}^{\prime}}\,
\exp\left(
\frac{T}{\stiff_{0}}\,\,
{\cal G}_{dd^{\prime}}({\bf x}-{\bf x}^{\prime})\,
k_{\rtr d}\cdot k_{\rtr d^{\prime}}^{\prime}
\right)
\nonumber\\ 
&=&
\widetilde{\cw}(k_{\rtr})\,
\widetilde{\cw}(k_{\rtr}^{\prime})\,
\delta_{{\bf k}_{\rm tot},{\bf k}_{\rm tot}^{\prime}}
\int\frac{d{\bf x}}{V}\,e^{ i{\bf k}_{\rm tot}\cdot{\bf x}}\,
\exp\left(
\frac{T}{\stiff_{0}}\,\,
{\cal G}_{dd^{\prime}}({\bf x})\,
k_{\rtr d}\cdot k_{\rtr d^{\prime}}^{\prime}
\right),
\label{eq:FTcorrel}
\end{eqnarray}
where $\widetilde{\cw}(k_{\rtr})$ is the 
fluctuation-renormalized order parameter given in 
Eqs.~(\ref{eq:rendistdef}) and (\ref{eq:shiftindistribution}).  
Next, in Eq.~(\ref{eq:CorrelMeaning}) we perform the integration 
over the center of mass of ${\bf r}_{1}$ and ${\bf r}_{2}$, and 
equate the resulting form of the correlator to the form given in 
Eq.~(\ref{eq:FTcorrel}), having dropped the hats on the dummy 
variables, thus arriving at a formula obeyed by the distribution 
$\cpd$: 
\begin{eqnarray}
&&
Q^{2}\,\delta_{{\bf k}_{1{\rm tot}},{\bf k}_{2{\rm tot}}}
\int d{\bf x}\,
     d{\Delta r}_{11}\,d{\Delta r}_{22}\,d{\Delta r}_{12}\,
\cpd({\bf x},{\Delta r}_{11},{\Delta r}_{22},{\Delta r}_{12})\,
e^{i{\bf k}_{1{\rm tot}}\cdot{\bf x}}
\nonumber\\ &&\qquad\qquad\qquad\qquad\times
e^{-\frac{1}{2}{\Delta r}_{11d_{1}d_{2}}
    \sum_{\alpha=0}^{n}k_{1d_{1}}^{\alpha}k_{1d_{2}}^{\alpha}}\,
e^{-\frac{1}{2}{\Delta r}_{22d_{1}d_{2}}
    \sum_{\alpha=0}^{n}k_{2d_{1}}^{\alpha}k_{2d_{2}}^{\alpha}}\,
e^{            {\Delta r}_{12d_{1}d_{2}}
    \sum_{\alpha=0}^{n}k_{1d_{1}}^{\alpha}k_{2d_{2}}^{\alpha}}
\nonumber\\ &&\qquad\qquad
=
\widetilde{\cw}(k_{1\rtr})\,
\widetilde{\cw}(k_{2\rtr})\,
\delta_{{\bf k}_{1{\rm tot}},{\bf k}_{2{\rm tot}}}
\int\frac{d{\bf x}}{V}\,e^{ i{\bf k}_{1{\rm tot}}\cdot{\bf x}}\,
\exp\left(
\frac{T}{\stiff_{0}}\,\,
{\cal G}_{d_{1}d_{2}}({\bf x})\,
k_{1\rtr d_{1}}\cdot k_{2\rtr d_{2}}
\right). 
\label{eq:ProToCol}
\end{eqnarray}
By solving this equation for 
$\cpd({\bf x},{\Delta r}_{11},{\Delta r}_{22},{\Delta r}_{12})$ 
one learns that it has weight only at values of 
$\big({\Delta r}_{11},{\Delta r}_{22},{\Delta r}_{12}\big)$
of the form 
$\big(
\xi_{1}^{2}\openone,\xi_{1}^{2}\openone,
({T}/{\stiff_{0}}){\cal G}_{d_{1}d_{2}}({\bf y})\big)$ 
for some values of the parameters 
$\big(\xi_{1},\xi_{1},{\bf y}\big)$, 
where $\openone$ is the identity in $D$-dimensional cartesian space. 
Anticipating this, it is convenient to introduce the following 
parametrization in terms of a reduced distribution 
$\cpd({\bf x},\xi_{1}^{2},\xi_{2}^{2},{\bf y})$: 
\begin{eqnarray}
\cpd({\bf x},{\Delta r}_{11},{\Delta r}_{22},{\Delta r}_{12})
&=&
\int
d\xi_{1}^{2}\,\wt{\dsl}\left(\xi_{1}^{2}\right)\,
d\xi_{2}^{2}\,\wt{\dsl}\left(\xi_{2}^{2}\right)\,
\delta\big({\Delta r}_{11}-\xi_{1}^{2}\openone\big)\,
\delta\big({\Delta r}_{22}-\xi_{2}^{2}\openone\big)
\nonumber\\
&&
\qquad\qquad\qquad
\qquad\qquad\qquad
\times
\int\frac{{d{\bf y}}}{V}\,
\delta\big({\Delta r}_{12}-({T}/{\stiff_{0}})\,
        {\cal G}({\bf y})\big) 
\,\cpd({\bf x},\xi_{1}^{2},\xi_{2}^{2},{\bf y}). 
\label{eq:parametrize}
\end{eqnarray}
Note that $\cpd({\bf x},\xi_{1}^{2},\xi_{2}^{2},{\bf y})$ 
only becomes a true distribution when an appropriate Jacobian 
factor associated with the ${\bf y}$ dependence is introduced; 
nevertheless we shall continue to refer to it as a distribution. 
As shown in App.~\ref{sec:CoToDist}, Eq.~(\ref{eq:ProToCol}) 
can be solved for the reduced probability distribution; 
the solution is given in Eq.~(\ref{eq:UnFourier}).
Thus one finds for the full distribution: 
\begin{eqnarray}
&&
\cpd({\bf x},{\Delta r}_{11},{\Delta r}_{22},{\Delta r}_{12})\,
=\int
d\xi_{1}^{2}\,\wt{\dsl}\left(\xi_{1}^{2}\right)\,
d\xi_{2}^{2}\,\wt{\dsl}\left(\xi_{2}^{2}\right)\,
\delta\big({\Delta r}_{11}-\xi_{1}^{2}\openone\big)\,
\delta\big({\Delta r}_{22}-\xi_{2}^{2}\openone\big)\,
\nonumber\\ && 
\qquad\quad\times
\int\frac{{d{\bf y}}}{V}
\left\{
\int\dbar{\bf q}\,
e^{-i{\bf q}\cdot({\bf x}-{\bf y})}\,
\exp\left(
 \frac{1}{2}\big(\xi_{1}^{2}+\xi_{2}^{2}\big)\,q^{2}
-{\Delta r}_{12d_{1}d_{2}}\,q_{d_{1}}\,q_{d_{2}}
    \right)
\right\}
\delta\big({\Delta r}_{12}-({T}/{\stiff_{0}})
        {\cal G}({\bf y})\big). 
\label{eq:RepeatUnFourier}
\end{eqnarray}
This has the structure of a source term, associated with the range 
of values of ${\cal G}$, convoluted with an appropriate 
\lq\lq propagator\rlap.\rq\rq

It should be borne in mind that, in this solution for the full 
distribution, Eq.~(\ref{eq:RepeatUnFourier}), the integration 
over wave vectors ${\bf q}$ is subject to the usual cut-offs, 
i.e., those featuring in Eq.~(\ref{eq:greenKS}).  Thus, even 
though the exponent in the factor 
$\exp\left(\frac{1}{2}\big(\xi_{1}^{2}+\xi_{2}^{2}\big)\,q^{2}
-{\Delta r}_{12d_{1}d_{2}}\,q_{d_{1}}\,q_{d_{2}}\right)$
{\it grows\/} at large ${\bf q}$ (i.e.~the factor is {\it not\/} a 
decaying Gaussian factor), the large-wave-vector cut-off protects 
against divergence.  In fact, if we focus on the distribution $\cpd$ 
at separations ${\bf x}$ that are large compared with the cut-off 
$\coshort$ then for values of $\xi_{1}$ and $\xi_{2}$ with 
appreciable weight the aforementioned exponent is small, and we 
may expand to obtain 
\begin{subequations}
\begin{eqnarray}
&&
\cpd({\bf x},{\Delta r}_{11},{\Delta r}_{22},{\Delta r}_{12})\,
=
\int
d\xi_{1}^{2}\,\wt{\dsl}\left(\xi_{1}^{2}\right)\,
d\xi_{2}^{2}\,\wt{\dsl}\left(\xi_{2}^{2}\right)\,
\delta\big({\Delta r}_{11}-\xi_{1}^{2}\openone\big)\,
\delta\big({\Delta r}_{22}-\xi_{2}^{2}\openone\big)\,
\nonumber
\\ 
&&
\qquad\qquad\qquad
\exp\left(
-\frac{1}{2}\big(\xi_{1}^{2}+\xi_{2}^{2}\big)\,\nabla_{\bf x}^{2}
+{\Delta r}_{12d_{1}d_{2}}\,
  \partial_{d_{1}}  
  \partial_{d_{2}}  
    \right)
\int\frac{{d{\bf y}}}{V}\,
\tilde{\delta}({\bf x}-{\bf y})\,
\delta\big({\Delta r}_{12}-({T}/{\stiff_{0}})
        {\cal G}({\bf y})\big)
\\
&&
\qquad
\approx\int
d\xi_{1}^{2}\,\wt{\dsl}\left(\xi_{1}^{2}\right)\,
d\xi_{2}^{2}\,\wt{\dsl}\left(\xi_{2}^{2}\right)\,
\delta\big({\Delta r}_{11}-\xi_{1}^{2}\openone\big)\,
\delta\big({\Delta r}_{22}-\xi_{2}^{2}\openone\big)\,
\nonumber\\ && 
\qquad\qquad\qquad
\times
\frac{1}{V}
\exp\left(
-\frac{1}{2}\big(\xi_{1}^{2}+\xi_{2}^{2}\big)\,\nabla_{\bf x}^{2}
+{\Delta r}_{12d_{1}d_{2}}\,
  \partial_{d_{1}}  
  \partial_{d_{2}}  
    \right)\,
\delta\big({\Delta r}_{12}-({T}/{\stiff_{0}})
        {\cal G}({\bf x})\big) 
\label{eq:ExpandUnFourier}
\\
&&
\qquad
\approx\int
d\xi_{1}^{2}\,\wt{\dsl}\left(\xi_{1}^{2}\right)\,
d\xi_{2}^{2}\,\wt{\dsl}\left(\xi_{2}^{2}\right)\,
\delta\big({\Delta r}_{11}-\xi_{1}^{2}\openone\big)\,
\delta\big({\Delta r}_{22}-\xi_{2}^{2}\openone\big)\,
\nonumber\\ 
&& 
\qquad\qquad\qquad
\times
\frac{1}{V}
\left\{
1-\frac{1}{2}\big(\xi_{1}^{2}+\xi_{2}^{2}\big)\,\nabla_{\bf x}^{2}
+{\Delta r}_{12d_{1}d_{2}}\,
  \partial_{d_{1}}  
  \partial_{d_{2}} 
   \right\}
\delta\big({\Delta r}_{12}-({T}/{\stiff_{0}})
        {\cal G}({\bf x})\big).
\label{eq:NewExpandUnFourier}
\end{eqnarray}
\end{subequations}
Here, $\partial_{d}\equiv\partial/\partial x_{d}$, and $\tilde{\delta}$ 
indicates the smoothed delta function resulting from the application of 
the wave-vector cut-off to the Fourier integration that yields it.  We 
have, however, proceded to the second and third lines without explicitly 
indicating the effect of this smoothing, viz., the replacement of the 
factor 
$\delta\big({\Delta r}_{12}-({T}/{\stiff_{0}}){\cal G}({\bf x})\big)$ 
the the same quantity {\it smeared\/} over a region around the point 
${\bf x}$ having linear dimension of order the short-distance cut-off.


\subsection{Two- (and higher-) field correlators 
as distributions of particle correlations}
Let us pause to revisit the interpretation of the two-field correlator.  
To do this, we introduce the disorder-averaged distribution 
${\cal M}_{2}$ of two-particle correlators ${\cal C}$, viz., 
\begin{equation}
{\cal M}_{2}[{\cal C}]
\!\equiv\!\!
\Big[\!
J^{-2}\!\!
\sum_{j_{1},j_{2}=1}^{J}
\prod_{{\bf k}_{1},{\bf k}_{2}\ne{\bf 0}}
\!\!\!\!
\delta_{\rm c}
\big(
{\cal C}_{{\bf k}_{1}{\bf k}_{2}}
\!-\!\langle
e^{i{\bf k}_{1}\cdot{\bf R}_{j_{1}}+
   i{\bf k}_{2}\cdot{\bf R}_{j_{2}}}
\rangle
\big)
\Big], 
\label{eq:disdistwo}
\end{equation}
in which $\delta_{\rm c}(x+iy)\equiv\delta(x)\,\delta(y)$.  
We can then observe that the two-field correlator, 
Eq.~(\ref{eq:repint}), can be expressed as a suitable moment 
of ${\cal M}_{2}$: 
\begin{equation}
\big\langle\op(k_{1})\,\op(k_{2})\big\rangle=
\int{\cal D}{\cal C}\,{\cal M}_{2}[{\cal C}]\,
\prod_{\alpha=0}^{n}
{\cal C}_{{\bf k}_{1}^{\alpha}{\bf k}_{2}^{\alpha}}\,.
\label{eq:disdistwomoment}
\end{equation}
Note that we have appealed to replica permutation symmetry 
in order to reinstate the dependence on the zeroth-replica 
wave vectors.  

%
%


It is straightforward to extend this discussion to the general 
case of $r$-particle correlators 
$\langle
e^{i{\bf k}_{1}\cdot{\bf R}_{j_{1}}
  +i{\bf k}_{1}\cdot{\bf R}_{j_{1}}
  +\cdots
  +i{\bf k}_{r}\cdot{\bf R}_{j_{r}}}\rangle$, 
for which the distribution ${\cal M}_{r}$ is given by 
\begin{equation}
{\cal M}_{r}[{\cal C}]
\equiv
\Big[
\frac{1}{J^{r}}
\sum_{j_{1},\ldots,j_{r}=1}^{J}
\prod_{{\bf k}_{1},{\bf k}_{2},\ldots,{\bf k}_{r}\ne{\bf 0}}
\delta_{\rm c}
\big(
{\cal C}_{{\bf k}_{1}{\bf k}_{2}\cdots{\bf k}_{r}}
-\langle
e^{i{\bf k}_{1}\cdot{\bf R}_{j_{1}}
  +i{\bf k}_{1}\cdot{\bf R}_{j_{1}}
  +\cdots
  +i{\bf k}_{r}\cdot{\bf R}_{j_{r}}}
\rangle
\big)
\Big], 
\end{equation}
Observe that the $r$-field correlator can also be expressed 
as a suitable moment: 
\begin{equation}
\big\langle\op(k_{1})\,\op(k_{2})
      \cdots\op(k_{r})\big\rangle
=\int{\cal D}{\cal C}\,{\cal M}_{r}[{\cal C}]\,
\prod_{\alpha=0}^{n}
{\cal C}_{{\bf k}_{1}^{\alpha}
          {\bf k}_{2}^{\alpha}\cdots
          {\bf k}_{r}^{\alpha}}\,\,.
\label{eq:disdisRmoment}
\end{equation}
Again we have appealed to replica permutation symmetry in order 
to reinstate the dependence on the zeroth-replica wave vectors. 
The distributions ${\cal M}_{r}$ are natural generalizations 
of the distribution of local density fluctuations, explored, 
e.g.~in Ref.~\cite{PengCastillo1998}.
\end{widetext}

\section{Concluding remarks}
\label{sec:conclusions}
In this Paper we have identified the long wave-length, low energy 
Goldstone-type fluctuations of the amorphous solid state, and 
investigated their physical consequences.  
By constructing an effective free energy governing these fluctuations, 
we have determined the elastic properties of the amorphous solid, 
including its static shear modulus which, we have re-confirmed, 
vanishes as the third power of the amount by which the constraint 
density exceeds its critical value (at the classical level).
We have also analyzed the effect of these fluctuations on the amorphous 
solid order parameter, finding that, in spatial dimensions greater 
than two, they induce a simple, rigid shift of the distribution of 
(squared) localization lengths.  
In addition, we have explored the properties of the order-parameter 
correlations in the amorphous solid state, establishing their physical 
content in terms of a joint probability distribution characterizing 
pairs of localized particles.  
Moreover, we have computed the corresponding correlator induced by 
Goldstone-type fluctuations and, hence, obtained a specific formula 
for this joint probability distribution. 

We have paid particular attention to systems of spatial dimension 
two.  In this setting we have shown that fluctuations restore the 
symmetries broken spontaneously at the classical level, particle 
localization is destroyed, the order parameter is driven to zero, 
and order-parameter correlations decay as a power-law in the 
separation between points in the sample.  The state is a 
{\it quasi-amorphous-solid\/} state, inasmuch as it possesses 
algebraically-decaying correlations and rigidity.  

Our work can be extended in several directions.  We have focussed here
on the critical behavior of the shear modulus. It is also of interest
to study elasticity in the strongly crosslinked limit, in which
(geometrical and thermal) fluctuations are less important and mean-field
theory should even be quantitatively correct. Whereas for long-chain 
macromolecules this limit is difficult to achieve, it may well be 
possible for Brownian particles, i.e., networks that are built from 
small units, in the simplest case just monomers~\cite{ref:KWZbrown}.
Even more interesting is a generalization to 
nonlinear elasticity.  Rubber can withstand very large 
deformations---up to $1000\mypercent$---thus allowing large 
amounts of elastic energy to be stored in the system.  
Our approach can be generalized to include nonlinear terms in 
the strain tensor, which will arise from two sources: 
the expansion of the Goldstone type fluctuations~(\ref{eq:GDk}) 
and higher-order nonlinearities in the Landau-Wilson free 
energy~(\ref{eq:Landaueffective}). Work along these lines is in progress. 

\bigskip\noindent{\it Acknowledgments\/}:
Certain ideas concerning the nature of the vulcanization transition 
in two spatial dimensions, developed in detail in the present Paper, 
originated in enlightening discussions involving 
Horacio E.~Castillo and Weiqun Peng; 
see Refs.~\cite{GoldbartTrieste2000} and~\cite{Peng+Goldbart2000}.  
It is a pleasure to acknowledge these discussions, as well as 
enlightening discussions with Matthew P.~A.~Fisher, Dominique Toublan 
and Michael Stone. 
We thank for their hospitality
the Department of Physics at the University of Colorado--Boulder (PMG)
and the Kavli Institute for Theoretical Physics at the University of
Calfornia--Santa Barbara (PMG and AZ), where some of the work reported
here was undertaken. 
This work was supported in part by the National Science Foundation 
under grants 
NSF DMR02-05858 (PMG, SM), 
NSF  PH99-07949 (PMG, AZ), 
and and by the DFG through SFB~602 and Grant No.~Zi 209/6-1 (AZ).

\appendix

\section{From Landau-Wilson hamiltonian to elastic free energy}
\label{sec:LWHtoEFE}
In this appendix we show how to get from the Landau-Wilson 
effective hamiltonian ${\cal S}_{\op}$, Eq.~(\ref{eq:Landaueffective}), 
to the elastic free energy~(\ref{eq:replicastiffness}).  Specifically, 
we compute the increase ${\cal S}_{u}$ in ${\cal S}_{\op}$ when the 
classical value $\op_{\rm cl}$, Eqs.~(\ref{eq:classicalstate}) and 
(\ref{eq:CSelements}), is replaced by Goldstone-disorted classical 
state~(\ref{eq:GDk}), parametrized by $u_{\rtr}({\bf x})$.  There are 
three terms in~(\ref{eq:Landaueffective}) to be computed. Two are 
\lq\lq potential\rlap,\rq\rq\ terms, which we shall see to have no 
dependence on $u_{\rtr}({\bf x})$; the third is a 
\lq\lq gradient\rq\rq\ term, and this is the origin of the 
dependence of ${\cal S}_{u}$ on $u_{\rtr}({\bf x})$.

Before focusing on any individual terms, we note that we can express 
summations over higher-replica sector wave vectors as unrestricted 
summations, less lower-replica sector contributions; e.g., 
\begin{equation}
\sum\nolimits_{k\in\hrs}=
\sum\nolimits_{k}-\sum\nolimits_{k\in\ors}-\sum\nolimits_{k\in\zrs}.
\end{equation}
Here, the two components of the lower-replica sector, viz., 
the one- and zero-replica sectors, are respectively denoted 
$\ors$ and $\zrs$. 

\begin{widetext}
Applying this sector decomposition to the first term in 
Eq.~(\ref{eq:Landaueffective}), and recognizing that there is no 
contribution from the one-replica sector and that the zero-replica 
sector contribution is simple, we have 
\begin{equation}
\sum\nolimits_{k\in\hrs}
\vert\op(k)\vert^{2}
=\left(  \sum\nolimits_{k}
        -\sum\nolimits_{k\in\ors}
        -\sum\nolimits_{k\in\zrs}
              \right)
\vert\op(k)\vert^{2}
=\sum\nolimits_{k}
\vert\op(k)\vert^{2}-0-Q^{2}.
\end{equation}
Recall that we are concerned with the increase in ${\cal S}_{\op}$ 
due to the Goldstone distortion.  Evidently, of the contributions 
considered so far only the unrestricted sum has the possibility of 
being sensitive to the distortion.  However, as we shall now see, 
not even this contribution has such sensitivity: 
\begin{eqnarray}
&&
\frac{1}{V^{n}}\sum_{k}\vert\op(k)\vert^{2}
=V\int\dbar k_{\rlo}\,\dbar k_{\rtr}
\int\frac{d{\bf x}_{1}}{V}
    \frac{d{\bf x}_{2}}{V}\,
e^{-i{\bf k}_{\rm tot}\cdot{\bf x}_{1}
   +i{\bf k}_{\rm tot}\cdot{\bf x}_{2}}
e^{-ik_{\rtr}\cdot u_{\rtr}({\bf x}_{1})
   +ik_{\rtr}\cdot u_{\rtr}({\bf x}_{2})}
   \,\cw(k_{\rtr})^{2}
\nonumber\\ &&\quad
=V\int\frac{d{\bf x}_{1}}{V}
      \frac{d{\bf x}_{2}}{V}
\int\dbar k_{\rtr}\,\cw(k_{\rtr})^{2}\,
e^{-ik_{\rtr}\cdot\left(
        u_{\rtr}({\bf x}_{1})-u_{\rtr}({\bf x}_{2})
                  \right)}
\int\dbar k_{\rlo}\,
e^{-i{\bf k}_{\rm tot}\cdot
        \left({\bf x}_{1}-{\bf x}_{2}\right)}
\nonumber\\ &&\quad
=V\int\frac{d{\bf x}_{1}}{V}
      \frac{d{\bf x}_{2}}{V}
\int\dbar k_{\rtr}\,\cw(k_{\rtr})^{2}\,
e^{-ik_{\rtr}\cdot\left(
        u_{\rtr}({\bf x}_{1})-u_{\rtr}({\bf x}_{2})
                  \right)}
(1+n)^{-\frac{D}{2}}\,\delta({\bf x}_{1}-{\bf x}_{2})
=(1+n)^{-\frac{D}{2}}\int\dbar k_{\rtr}\,\cw(k_{\rtr})^{2},
\end{eqnarray}
independent of $u_{\rtr}({\bf x})$.  Note that here and elsewhere 
in the present Apendix we shall anticipate the taking of the replica 
limit by omiting factors of $V^{n}$. 

Turning to the third term in~(\ref{eq:Landaueffective}), the 
nonlinearity, and handling the constraints on the summation with 
care, we are faced with the term 
\def\apshifter{\quad}
\begin{eqnarray}
&&
V\sum_{k_{1},k_{2},k_{3}}
\delta_{k_{1}+k_{2}+k_{3},0}\,\,
\op(k_{1})\,\op(k_{2})\,\op(k_{3})
=
\int\dbar k_{1}\,\dbar k_{2}\,\dbar k_{3}\,
\int{dy}\,e^{-iy\cdot\left(k_{1}+k_{2}+k_{3}\right)}
\nonumber\\ &&\qquad\qquad\times
\int{d{\bf x}_{1}}\,
    {d{\bf x}_{2}}\,
    {d{\bf x}_{3}}\,
    e^{ i{\bf k}_{1\rm tot}\cdot{\bf x}_{1}
       +i{\bf k}_{2\rm tot}\cdot{\bf x}_{2}
       +i{\bf k}_{3\rm tot}\cdot{\bf x}_{3}}
e^{ ik_{1\rtr}\cdot u_{\rtr}({\bf x}_{1})
   +ik_{2\rtr}\cdot u_{\rtr}({\bf x}_{2})
   +ik_{3\rtr}\cdot u_{\rtr}({\bf x}_{3})}\,
   \cw(k_{1\rtr})\,\cw(k_{2\rtr})\,\cw(k_{3\rtr})
\nonumber\\ &&\apshifter
=
\int\dbar k_{1\rlo}\,\dbar k_{2\rlo}\,\dbar k_{3\rlo}\,
\int\frac{dy_{\rlo}}{V}
e^{-iy_{\rlo}\cdot\left(k_{1\rlo}+k_{2\rlo}+k_{3\rlo}\right)}
\int\dbar k_{1\rtr}\,\dbar k_{2\rtr}\,\dbar k_{3\rtr}\,
\int{dy_{\rtr}}
e^{-iy_{\rtr}\cdot\left(k_{1\rtr}+k_{2\rtr}+k_{3\rtr}\right)}
\nonumber\\ &&\qquad\qquad\times
\int{d{\bf x}_{1}}\,
    {d{\bf x}_{2}}\,
    {d{\bf x}_{3}}\,
    e^{ i{\bf k}_{1\rm tot}\cdot{\bf x}_{1}
       +i{\bf k}_{2\rm tot}\cdot{\bf x}_{2}
       +i{\bf k}_{3\rm tot}\cdot{\bf x}_{3}}
e^{ ik_{1\rtr}\cdot u_{\rtr}({\bf x}_{1})
   +ik_{2\rtr}\cdot u_{\rtr}({\bf x}_{2})
   +ik_{3\rtr}\cdot u_{\rtr}({\bf x}_{3})}\,
   \cw(k_{1\rtr})\,\cw(k_{2\rtr})\,\cw(k_{3\rtr})
\nonumber\\ &&\apshifter
=
(1+n)^{D}
\int\dbar{\bf k}_{1}\,\dbar{\bf k}_{2}\,\dbar{\bf k}_{3}\,
\int\frac{d{\bf y}}{V}\,
e^{-i{\bf y}\cdot\left({\bf k}_{1}+{\bf k}_{2}+{\bf k}_{3}\right)}
\int\dbar k_{1\rtr}\,\dbar k_{2\rtr}\,\dbar k_{3\rtr}\,
\int{dy_{\rtr}}\,
e^{-iy_{\rtr}\cdot\left(k_{1\rtr}+k_{2\rtr}+k_{3\rtr}\right)}
\nonumber\\ &&\qquad\qquad\times
\int{d{\bf x}_{1}}\,
    {d{\bf x}_{2}}\,
    {d{\bf x}_{3}}\,
    e^{ i{\bf k}_{1}\cdot{\bf x}_{1}
       +i{\bf k}_{2}\cdot{\bf x}_{2}
       +i{\bf k}_{3}\cdot{\bf x}_{3}}
e^{ ik_{1\rtr}\cdot u_{\rtr}({\bf x}_{1})
   +ik_{2\rtr}\cdot u_{\rtr}({\bf x}_{2})
   +ik_{3\rtr}\cdot u_{\rtr}({\bf x}_{3})}\,
   \cw(k_{1\rtr})\,\cw(k_{2\rtr})\,\cw(k_{3\rtr})
\nonumber\\ &&\apshifter
=
(1+n)^{D}
\int\dbar k_{1\rtr}\,\dbar k_{2\rtr}\,\dbar k_{3\rtr}\,\,
{\delta_{k_{1\rtr}+k_{2\rtr}+k_{3\rtr},0}}
\int\frac{d{\bf y}}{V}
\int{d{\bf x}_{1}}\,
    {d{\bf x}_{2}}\,
    {d{\bf x}_{3}}\,\,
\delta({\bf x}_{1}-{\bf y})\,
\delta({\bf x}_{2}-{\bf y})\,
\delta({\bf x}_{3}-{\bf y})\,
\nonumber\\ &&\qquad\qquad\times
e^{ ik_{1\rtr}\cdot u_{\rtr}({\bf x}_{1})
   +ik_{2\rtr}\cdot u_{\rtr}({\bf x}_{2})
   +ik_{3\rtr}\cdot u_{\rtr}({\bf x}_{3})}\,
   \cw(k_{1\rtr})\,\cw(k_{2\rtr})\,\cw(k_{3\rtr})
\nonumber\\ &&\apshifter
=
{(1+n)^{D}}
\int\dbar k_{1\rtr}\,\dbar k_{2\rtr}\,\dbar k_{3\rtr}\,\,
\delta_{k_{1\rtr}+k_{2\rtr}+k_{3\rtr},0}
\int \frac{d{\bf y}}{V}\,
e^{i\left(k_{1\rtr}+k_{2\rtr}+k_{3\rtr}\right)
                \cdot u_{\rtr}({\bf y})}\,
   \cw(k_{1\rtr})\,\cw(k_{2\rtr})\,\cw(k_{3\rtr})
\nonumber\\ &&\apshifter
=
{(1+n)^{D}}
\int\dbar k_{1\rtr}\,\dbar k_{2\rtr}\,\dbar k_{3\rtr}\,\,
\delta_{k_{1\rtr}+k_{2\rtr}+k_{3\rtr},0}\,
\cw(k_{1\rtr})\,\cw(k_{2\rtr})\,\cw(k_{3\rtr}).
\end{eqnarray}
This is independent of $u_{\rtr}({\bf x})$. 
\end{widetext}

Turning to the second term in~(\ref{eq:Landaueffective}), 
the gradient term, we have 
\begin{eqnarray}
&&
\frac{1}{V^{n}}\sum_{k}{k\cdot k}\,
\vert\,\op(k)\vert^{2}
=V\int\dbar k_{\rlo}\,\,\dbar k_{\rtr}
\left(k_{\rlo}\cdot k_{\rlo}+k_{\rtr}\cdot k_{\rtr}\right)
\nonumber\\ &&\qquad\qquad\times
\int\frac{d{\bf x}_{1}}{V}
    \frac{d{\bf x}_{2}}{V}\,
e^{-i{\bf k}_{\rm tot}\cdot{\bf x}_{1}
   +i{\bf k}_{\rm tot}\cdot{\bf x}_{2}}
\nonumber\\ &&\qquad\qquad\times
e^{-ik_{\rtr}\cdot u_{\rtr}({\bf x}_{1})
   +ik_{\rtr}\cdot u_{\rtr}({\bf x}_{2})}
   \,\cw(k_{\rtr})^{2}.
\end{eqnarray}
Of the two contributions airising from this term, 
from $k_{\rlo}^{2}$ and 
from $k_{\rtr}^{2}$, 
the latter has no $u_{\rtr}({\bf x})$ dependence.  
This follows via the mechanism that we saw for the first term, viz., 
the development of a factor of $\delta({\bf x}_{1}-{\bf x}_{2})$.
As for the former contribution, to evaluate it we replace 
$k_{\rlo}\cdot k_{\rlo}$ by 
$(1+n)^{-1}{\bf k}_{\rm tot}\cdot{\bf k}_{\rm tot}$ 
and generate this factor via suitable derivatives: 
\begin{eqnarray}
&&
\frac{1}{V^{n}}\sum_{k}{k\cdot k}\,
\vert\op(k)\vert^{2}
=V\int\dbar k_{\rlo}\,\,\dbar k_{\rtr}\,
k_{\rlo}\cdot k_{\rlo}
\nonumber\\ &&\qquad\qquad\times
\int\frac{d{\bf x}_{1}}{V}
    \frac{d{\bf x}_{2}}{V}\,
e^{-i{\bf k}_{\rm tot}\cdot{\bf x}_{1}
   +i{\bf k}_{\rm tot}\cdot{\bf x}_{2}}
\nonumber\\ &&\qquad\qquad\times
e^{-ik_{\rtr}\cdot u_{\rtr}({\bf x}_{1})
   +ik_{\rtr}\cdot u_{\rtr}({\bf x}_{2})}
   \,\cw(k_{\rtr})^{2}
\nonumber\\ &&\quad
=(1+n)^{-D/2}(1+n)^{-1}
V\int\dbar k_{\rtr}\,\,
     \dbar{\bf k}_{\rm tot}\,
\nonumber\\ &&\qquad\qquad\times
\int\frac{d{\bf x}_{1}}{V}
    \frac{d{\bf x}_{2}}{V}\,
\left(
\partial_{-i{\bf x}_{1}}
e^{-i{\bf k}_{\rm tot}\cdot{\bf x}_{1}}
\right)
\cdot
\left(
\partial_{i{\bf x}_{2}}
e^{i{\bf k}_{\rm tot}\cdot{\bf x}_{2}}
\right)
\nonumber\\ &&\qquad\qquad\times
e^{-ik_{\rtr}\cdot u_{\rtr}({\bf x}_{1})}\,
e^{ ik_{\rtr}\cdot u_{\rtr}({\bf x}_{2})}\,
   \,\cw(k_{\rtr})^{2}.
\end{eqnarray}
Next, we integrate by parts, 
once with respect to ${\bf x}_{1}$ and 
once with respect to ${\bf x}_{2}$, 
to transfer the derivatives to the exponentional factors 
containing $u_{\rtr}$, arriving at  
\begin{eqnarray}
&&(1+n)^{-1-D/2}V
\int\frac{d{\bf x}_{1}}{V}
    \frac{d{\bf x}_{2}}{V}
\nonumber\\ &&
\quad\times
\int\dbar{\bf k}_{\rm tot}\,\,
e^{-i{\bf k}_{\rm tot}\cdot\left({\bf x}_{1}-{\bf x}_{2}\right)}
\int\dbar k_{\rtr}\,\cw(k_{\rtr})^{2}
\nonumber\\ &&
\quad\times
\left(
\partial_{-i{\bf x}_{1}}
e^{-ik_{\rtr}\cdot u_{\rtr}({\bf x}_{1})}
\right)
\cdot
\left(
\partial_{i{\bf x}_{2}}
e^{ ik_{\rtr}\cdot u_{\rtr}({\bf x}_{2})}
\right).
\end{eqnarray}
Powers of $k_{\rlo}$ higher than two would, via the corresponding 
derivatives, produce higher gradients of $u_{\rtr}$, as well as 
nonlinearities involving lower-order derivatives. By performing the 
derivatives that we have here, as well as the ${\bf k}_{\rm tot}$ 
integration, which generates a factor $\delta({\bf x}_{1}-{\bf x}_{2})$ 
and allows us to integrate over, say, ${\bf x}_{2}$, we obtain
\begin{eqnarray}
&&(1+n)^{-1-D/2}V^{-1}
\int\dbar k_{\rtr}\,\cw(k_{\rtr})^{2}
\nonumber\\ &&
\quad\times
\int d{\bf x}
\left({k_{\rtr}\cdot
\partial_{{\bf x}}u_{\rtr}({\bf x})}\right)
\cdot
\left(k_{\rtr}\cdot
\partial_{{\bf x}}{u_{\rtr}({\bf x})}\right), 
\label{eq:needcontract}
\end{eqnarray}
where the scalar products inside the parentheses are over $nD$-component 
replica-transverse vectors whilst the outside scalar product is over 
$D$-component position vectors.  The next step is to observe that, 
owing to the rotationally invariant form of $\cw(k_{\rtr})$, 
Eq.~(\ref{eq:mfstate}), the $k_{\rtr}$ integration includes an 
isotropic average of two components of $k_{\rtr}$, and is therefore 
proportional to the identity in $nD$-dimensional replica-transverse 
space.  Thus, we arrive at the form
\begin{eqnarray}
&&\frac{(1+n)^{-1-D/2}}{nD}
\int\dbar k_{\rtr}\,
\cw(k_{\rtr})^{2}\,\,k_{\rtr}\cdot k_{\rtr}
\nonumber\\ &&
\qquad\qquad\qquad\times
\int \frac{d{\bf x}}{V}\,
\left(
\partial_{{\bf x}}u_{\rtr}({\bf x})
\partial_{{\bf x}}u_{\rtr}({\bf x})
\right), 
\end{eqnarray}
which involves the two types of scalar product mentioned beneath 
Eq.~(\ref{eq:needcontract}).  Reinstating the factor of $V\den$ 
from Eq.~(\ref{eq:Landaueffective}), we identify the stiffness 
divided by the temperature) $\stiff_{n}/T$ and, hence, arrive at 
Eqs.~(\ref{eq:elasticcouple}).

\section{Evaluating the stiffness
 (a.k.a.~shear modulus or rigidity)}
\label{sec:ClassShear}
In this appendix we display the main steps for obtaining the 
stiffness $\stiff_{0}$, Eqs.~(\ref{eq:elasticstiffness}), by 
evaluating the formula for 
$\stiff_{n}$, given in Eq.~(\ref{eq:replicastiffness}) and 
derived in App.~\ref{sec:LWHtoEFE}, in terms of the elements of the 
classical state, Eqs.~(\ref{eq:CSelements}).  This involves the 
evaluation of the RHS of Eq.~(\ref{eq:replicastiffness}), which 
proceeds as follows: 
\def\shrsp{\!\!\!\!\!}
\begin{eqnarray}
&&\shrsp
\int\dbar k_{\rtr}\,k_{\rtr}^{2}\,\cw(k_{\rtr})^{2}
\nonumber\\
&&=
Q^{2}\int\dbar k_{\rtr}\,k_{\rtr}^{2}
\int_{0}^{\infty}\shrsp
d\xi^{2}\,\dsl\big(\xi^{2}\big)\,
d\xi^{\prime 2}\,\dsl\big(\xi^{\prime 2}\big)\,
e^{-\left(\xi^{2}+\xi^{\prime 2}\right)k_{\rtr}^{2}/2}
\nonumber\\
&&=
Q^{2}\int_{0}^{\infty}\shrsp
d\xi^{2}\,\dsl\big(\xi^{2}\big)\,
d\xi^{\prime 2}\,\dsl\big(\xi^{\prime 2}\big)
\int\dbar k_{\rtr}\,k_{\rtr}^{2}
e^{-\left(\xi^{2}+\xi^{\prime 2}\right)k_{\rtr}^{2}/2}
\nonumber\\
&&=
Q^{2}\int_{0}^{\infty}\shrsp
d\xi^{2}\,\dsl\big(\xi^{2}\big)\,
d\xi^{\prime 2}\,\dsl\big(\xi^{\prime 2}\big)
\nonumber\\
&&\qquad\qquad\times
\partial_{-A/2}\big\vert_{A=\xi^{2}+\xi^{\prime 2}}
\int\dbar k_{\rtr}\,e^{-Ak_{\rtr}^{2}/2}
\nonumber\\
&&=
Q^{2}\int_{0}^{\infty}\shrsp
d\xi^{2}\,\dsl\big(\xi^{2}\big)\,
d\xi^{\prime 2}\,\dsl\big(\xi^{\prime 2}\big)
\nonumber\\
&&\qquad\qquad\times
\partial_{-A/2}\big\vert_{A=\xi^{2}+\xi^{\prime 2}}
(2\pi A)^{-nD/2}
\nonumber\\
&&=
Q^{2}\int_{0}^{\infty}\shrsp
d\xi^{2}\,\dsl\big(\xi^{2}\big)\,
d\xi^{\prime 2}\,\dsl\big(\xi^{\prime 2}\big)
\nonumber\\
&&\qquad\qquad\times
nD(2\pi A)^{-nD/2}A^{-1}
\big\vert_{A=\xi^{2}+\xi^{\prime 2}}
\nonumber\\
&&\mathrel{\mathop{\approx}^{n\to 0}}
nDQ^{2}\int_{0}^{\infty}\shrsp
d\xi^{2}\,\dsl\big(\xi^{2}\big)\,
d\xi^{\prime 2}\,\dsl\big(\xi^{\prime 2}\big)
\left(\xi^{2}+\xi^{\prime 2}\right)^{-1}
\end{eqnarray}

\section{Elastic Green function}
\label{sec:EGF}
The elastic Green function
${\cal G}_{dd^{\prime}}({\bf x}-{\bf x}^{\prime})$ 
featuring in Eq.~(\ref{eq:uuviawick}) arises via functional 
integration over the displacement field ${\bf u}({\bf x})$.  
This integration comprises fields configurations that are 
(i)~volume-preserving (at least to leading order in the gradient 
of the displacement field), and 
(ii)~have Fourier content only from wave-lengths lying between 
the short- and long-distance cut-offs $\coshort$ and $\colong$. 
As the weight in Eq.~(\ref{eq:uuviawick}) is gaussian with 
respect to ${\bf u}({\bf x})$, we have that 
$(T/\stiff_{0})\,{\cal G}_{dd^{\prime}}({\bf x}-{\bf x}^{\prime})$ 
is proportional to the inverse of the operator appearing 
sandwiched between two displacement fields in the exponent of 
the gaussian, 
\begin{equation}
\frac{\stiff_{0}}{T}
\int_{\cal V} d{\bf x}\,
\big(
\partial_{\bf x}{\bf u}\cdot
\partial_{\bf x}{\bf u}
\big),
\label{eq:sandwich}
\end{equation}
provided this inverse is the one associated with the Hilbert 
space of vector-field configurations contributing to the 
functional integral.  An application of the divergence 
theorem shows that, in addition to conditions~(i) and (ii), 
the Green function obeys the equation 
\begin{equation}
-\nabla_{x}^{2}\,
{\cal G}_{dd^{\prime}}({\bf x})=
\delta_{dd^{\prime}}\,
\delta({\bf x}), 
\label{eq:EqForG}
\end{equation}
in which the delta function is to be interpreted as the identity 
in the appropriate Hilbert space, mentioned above.  To determine 
${\cal G}_{dd^{\prime}}({\bf x})$ 
we express it in its Fourier representation: 
\begin{equation}
{\cal G}_{dd^{\prime}}({\bf x})=
\int\dbar{\bf k}\,
e^{-i{\bf k}\cdot{\bf x}}\,
{\cal G}_{dd^{\prime}}({\bf k}). 
\label{eq:UnkFrep}
\end{equation}
Next, we insert this representation into Eq.~(\ref{eq:EqForG}) 
to obtain 
\begin{eqnarray}
&&
-\nabla_{x}^{2}
\int\dbar{\bf k}\,
e^{-i{\bf k}\cdot{\bf x}}\,
{\cal G}_{dd^{\prime}}({\bf k})
=\int\dbar{\bf k}\,
k^{2}
e^{-i{\bf k}\cdot{\bf x}}\,
{\cal G}_{dd^{\prime}}({\bf k})
\nonumber\\
&&\qquad
=\int_{2\pi/\colong}^{2\pi/\coshort}
\dbar{\bf k}\,
e^{-i{\bf k}\cdot{\bf x}}
\big(\delta_{dd^{\prime}}-k^{-2}\,k_{d}\,k_{d^{\prime}}\big). 
\label{eq:SolveIt}
\end{eqnarray}
where 
$\cbshort\equiv\coshort/2\pi$ and 
$\cblong \equiv\colong /2\pi$. 
The final term, the integral representation of the appropriate 
delta function, accommodates restrictions~(i) and (ii).  Then, 
from the linear independence of the plane waves, we see that the 
Fourier integral representation of the Green function is given 
by Eq.~(\ref{eq:UnkFrep}), with the amplitude given by 
\begin{equation}
{\cal G}_{dd^{\prime}}({\bf k})=
\begin{cases}
\big(k^{2}\,\delta_{dd^{\prime}}-k_{d}\,k_{d^{\prime}}\big)/k^{4},  
&{\rm for}\,\,\,
\cblong^{-1}<k<\cbshort^{-1};
\\
\noalign{\medskip}
0,
&{\rm otherwise}; 
\end{cases}
\end{equation}
as given in Eq.~(\ref{eq:greenKS}).

Before computing the real-space form of this $D$-dimensional 
Green function, we evaluate it at argument ${\bf x}={\bf 0}$, 
as well as at small argument ($\vert{\bf x}\vert\ll\coshort$). 
At ${\bf x}={\bf 0}$ we have 
\begin{subequations}
\begin{eqnarray}
{\cal G}_{dd^{\prime}}({\bf x})\vert_{{\bf x}={\bf 0}}
&=&\int_{\rm h.c.o.}
\!\!\!\!\!\!\!\!
\dbar{\bf k}\,
\left({k^{2}\,\delta_{dd^{\prime}}-k_{d}\,k_{d^{\prime}}}\right)
{k^{-4}}
\\
&=&
\delta_{dd^{\prime}}\frac{D-1}{D}
\int_{\rm h.c.o.}
\!\!\!\!\!\!\!\!
\dbar{\bf k}\,k^{-2}
\\
&=&
\delta_{dd^{\prime}}\,\GatZ_{\! D}\,,
\\
\GatZ_{\! D}&\equiv&
\frac{D-1}{D}
\frac{\Sigma_{D}}{(2\pi)^{D}}
\int_{\cblong^{-1}}^{\cbshort^{-1}}
\!\!
dk\,k^{D-3}.
\label{eq:hcoDef}
\end{eqnarray}
\end{subequations} 
Here and elsewhere, the subscript h.c.o.~indicates that the 
integration is subject to the hard cut-off shown 
explicitly in Eq.~(\ref{eq:hcoDef}).
Evaluating the last integral for $D\ge 2$, and for 
$D>2$ retaining only the dominant contribution, gives 
$\GatZ_{\! D}$, as given in Eq.~(\ref{eq:casesofGzero}). 
 
Now generalizing to $\vert{\bf x}\vert$ small but nonzero, 
i.e., $\vert{\bf x}\vert\ll\coshort$, we have, 
by expanding the exponential in Eq.~(\ref{eq:UnkFrep}), 
\begin{eqnarray}
{\cal G}_{dd^{\prime}}({\bf x})
&\approx&
\delta_{dd^{\prime}}\,\GatZ_{\! D}
-\frac{1}{2}[(D+1)\delta_{dd^{\prime}}-2\hat{x}_{d}\,\hat{x}_{d^{\prime}}]
\,\vert{\bf x}\vert^{2}
\nonumber
\\
&&\quad\times
\frac{\Sigma_{D}}{(2\pi)^{D}}
\frac{1}{D(D+2)}
\int_{\cblong^{-1}}^{\cbshort^{-1}}\!\!
dk\,k^{D-1}
\\
&\approx&
\delta_{dd^{\prime}}\,\GatZ_{\! D}
-\frac{1}{2}[(D+1)\delta_{dd^{\prime}}-2\hat{x}_{d}\,\hat{x}_{d^{\prime}}]
\,\vert{\bf x}\vert^{2}
\nonumber
\\
&&\quad\times
\frac{\Sigma_{D}}{\coshort^{D}}
\frac{1}{D^{2}(D+2)}, 
\end{eqnarray}
where, again, we have retained only the dominant contribution. 
Specializing to $D=2$ and $D=3$ we find 
\begin{eqnarray}
&&
\!\!\!\!\!\!\!\!\!\!
\!\!\!\!\!\!\!\!\!\!
{\cal G}_{dd^{\prime}}^{(2)}({\bf x})
\approx
\delta_{dd^{\prime}}\,\GatZ_{\! 2}
-\frac{1}{64\pi}\frac{\vert{\bf x}\vert^{2}}{\cbshort^{2}}
[3\,\delta_{dd^{\prime}}-2\hat{x}_{d}\,\hat{x}_{d^{\prime}}], 
\label{eq:GshortTwo}
\\
&&
\!\!\!\!\!\!\!\!\!\!
\!\!\!\!\!\!\!\!\!\!
{\cal G}_{dd^{\prime}}^{(3)}({\bf x})
\approx
\delta_{dd^{\prime}}\,\GatZ_{\! 3}
-\frac{1}{45\pi\coshort}\frac{\vert{\bf x}\vert^{2}}{\cbshort^{2}}
[2\,\delta_{dd^{\prime}}-\hat{x}_{d}\,\hat{x}_{d^{\prime}}]. 
\label{eq:GshortThree}
\end{eqnarray}

We now compute the real-space form of this 
$D$-dimensional Green function, valid for 
arbitrary $\vert{\bf x}\vert$:   
\def\PGlocTS{\!\!}
\begin{eqnarray} 
{\cal G}_{dd^{\prime}}({\bf x})
\PGlocTS&=&\PGlocTS
\int\dbar{\bf k}\,e^{-i{\bf k}\cdot{\bf x}}\,
{\cal G}_{dd^{\prime}}({\bf k})
\\
\PGlocTS&=&\PGlocTS
\int_{\rm h.c.o.}
\!\!\!\!\!\!\!\!
\dbar{\bf k}\,e^{-i{\bf k}\cdot{\bf x}}\,
\left({k^{2}\,\delta_{dd^{\prime}}-k_{d}\,k_{d^{\prime}}}\right)
{k^{-4}}
\\
\PGlocTS&=&\PGlocTS
-\left(
\delta_{dd^{\prime}}\nabla^{2}-\partial_{d}\,\partial_{d^{\prime}}
\right)
{\cal H}({\bf x}),
\label{eq:HUse}
\\
{\cal H}({\bf x})
\PGlocTS&\equiv&\PGlocTS
\int_{\rm h.c.o.}
\!\!\!\!\!\!\!\!
\dbar{\bf k}\,
e^{-i{\bf k}\cdot{\bf x}}\,
{k^{-4}}
\\
\PGlocTS&=&\PGlocTS
\int_{\cblong^{-1}}^{\cbshort^{-1}}
\!\!
\dbar k\,k^{D-1}
\int\dbar^{D-1}{\hat{\bf k}}\,\,
\frac{e^{-i{\bf k}\cdot{\bf x}}}{k^{4}}.
\end{eqnarray}

Specializing to the case of $D=3$, and using spherical polar 
co-ordinates $(k,\theta,\varphi)$, we have 
\begin{eqnarray} 
{\cal H}^{(3)}({\bf x})
&=&
(2\pi)^{-3}
\int_{\cblong}^{\cbshort}\frac{dk}{k^{2}}
\int_{0}^{2\pi}\!\!\!\! d\varphi\!
\int_{0}^{\pi} \!\! d\theta\,\sin\theta\,
e^{-ik\vert{\bf x}\vert\cos\theta}
\nonumber\\ 
\quad
&=&
\frac{\vert{\bf x}\vert}{2\pi^{2}}
\int_{\vert{\bf x}\vert/\cblong}^{\vert{\bf x}\vert/\cbshort}
dz\,\frac{\sin z}{z^{3}}.
\end{eqnarray}
To control the potential divergence at small $z$ we add and 
subtract the small-$z$ behavior of $\sin z$, thus obtaining 
\begin{eqnarray} 
{\cal H}^{(3)}({\bf x})
&=&
\frac{\vert{\bf x}\vert}{2\pi^{2}}
\int_{\vert{\bf x}\vert/{\cblong}}^{\vert{\bf x}\vert/{\cbshort}}
dz\,\left(\frac{\sin z -z}{z^{3}}+\frac{1}{z^{2}}\right)
\\
&=&
\frac{\vert{\bf x}\vert}{2\pi^{2}}
 \Big({\rm Si}_{3}(\vert{\bf x}\vert/\cbshort)-
      {\rm Si}_{3}(\vert{\bf x}\vert/\cblong)\Big)
\nonumber\\
&&\qquad\qquad\qquad\qquad
+\frac{\cblong-\cbshort}{2\pi^{2}},
\label{eq:EvalThrH}
\\
{\rm Si}_{3}(t)
&\equiv&
\int_{0}^{t}dz\,\,\frac{\sin z -z}{z^{3}}, 
\label{eq:Hthree}
\end{eqnarray} 
where ${\rm Si}_{3}(z)$ is a generalized sine integral, 
which has asymptotic behavior  
\begin{equation}
{\rm Si}_{3}(t)
\approx
\begin{cases}
-{t}/{3!},&{\rm for}\,\,\,t\ll 1,
\\
-(\pi/4)-t^{-1},&{\rm for}\,\,\,t\gg 1.
\end{cases}
\end{equation}
Using this behavior to approximate 
${\cal H}^{(3)}({\bf x})$ in Eq.~(\ref{eq:EvalThrH}), 
and inserting the result into Eq.~(\ref{eq:HUse}), noting 
that the final term vanishes under differentiation, we obtain 
for $\coshort\ll\vert{\bf x}\vert\ll\colong$ 
\begin{eqnarray}
{\cal G}_{dd^{\prime}}^{(3)}({\bf x})
&\approx&
\left(
\delta_{dd^{\prime}}\nabla^{2}-\partial_{d}\,\partial_{d^{\prime}}
\right)
\frac{\vert{\bf x}\vert}{8\pi}
\\
&=&
\frac{1}{8\pi\,\vert{\bf x}\vert} 
\left(\delta_{dd^{\prime}}+
{\hat{x}}_{d}\,{\hat{x}}_{d^{\prime}}\right)
\end{eqnarray}

Now specializing to the case of $D=2$, and using plane polar 
co-ordinates $(k,\varphi)$, we have 
\begin{eqnarray} 
{\cal H}^{(2)}({\bf x})
&=&
(2\pi)^{-2}
\int_{\cblong^{-1}}^{\cbshort^{-1}}
\frac{dk}{k}
\int_{0}^{2\pi}\!\!\!\! d\varphi\,
e^{-ik\vert{\bf x}\vert\cos\varphi}
\nonumber\\ 
\quad
&=&
\frac{\vert{\bf x}\vert^{2}}{2\pi}
\int_{\vert{\bf x}\vert/\cblong}^{\vert{\bf x}\vert/\cbshort}
\frac{dz}{z^{3}}
\int_{0}^{2\pi}\!\!\!\! \dbar\varphi\,
e^{-iz\cos\varphi}.
\end{eqnarray}
To control the potential divergence at small $z$ we add and 
subtract the small-$z$ behavior of the integrand, thus obtaining 
\def\sfn{{\cal S}}
\begin{eqnarray} 
{\cal H}^{(2)}({\bf x})
&=&
\frac{\vert{\bf x}\vert^{2}}{2\pi}
\int_{\vert{\bf x}\vert/\cblong}^{\vert{\bf x}\vert/\cbshort}
\frac{dz}{z^{3}}\,
\int_{0}^{2\pi}\!\!\!\dbar\varphi
\left(
e^{-iz\cos\varphi}
-\left[1-\smquar z^{2}\right]
\right)
\nonumber \\
&&\qquad\qquad+
\frac{\vert{\bf x}\vert^{2}}{2\pi}
\int_{\vert{\bf x}\vert/\cblong}^{\vert{\bf x}\vert/\cbshort}
\frac{dz}{z^{3}}\,
\left(1-\smquar z^{2}\right)
\\
&=&
\frac{\vert{\bf x}\vert^{2}}{2\pi}
\Big(
\sfn(\vert{\bf x}\vert/\cbshort)-
\sfn(\vert{\bf x}\vert/\cblong)
\Big)
  +\frac{\cblong^{2}-\cbshort^{2}}{4\pi}
\nonumber \\
&&\qquad\qquad\qquad
-\frac{\vert{\bf x}\vert^{2}}{8\pi}
 \ln\left(\cblong/\cbshort\right), 
\label{eq:EvalTwoH}
\\
\sfn(t)&\equiv&
\int_{0}^{t}
\frac{dz}{z^{3}}\,
\int_{0}^{2\pi}\!\!\!
\dbar\varphi\,
\left(
e^{-iz\cos\varphi}
-\left[1-\smquar z^{2}\right]
\right).
\end{eqnarray}
Noting that $\sfn(t)$ has asymptotic behavior~\cite{REF:Stone}
\begin{equation}
\sfn(t)
\approx
\begin{cases}
\frac{1}{128}t^{2},&{\rm for}\,\,\,t\ll 1,
\\
\noalign{\medskip}
\frac{1}{4}\ln\big(t/2e^{1-\gamma}\big), &{\rm for}\,\,\,t\gg 1, 
\end{cases}
\end{equation}
where $\gamma$ ($=0.5772\ldots$) is the Euler-Mascheroni constant
and using this to approximate 
${\cal H}^{(2)}({\bf x})$ in Eq.~(\ref{eq:EvalTwoH}), 
and inserting the result into Eq.~(\ref{eq:HUse}), 
noting that the constant term vanishes under 
differentiation, we obtain for 
$\coshort\ll\vert{\bf x}\vert\ll\colong$ 
the result 
\begin{equation}
{\cal G}_{dd^{\prime}}^{(2)}({\bf x})
\approx
-\frac{1}{4\pi}\Big(
\delta_{dd^{\prime}}
\ln\left(e^{\gamma+\frac{1}{2}}\,\vert{\bf x}\vert/2\cblong\right)
-{\hat{x}}_{d}\,{\hat{x}}_{d^{\prime}}\Big).
\label{eq:EGFtwoDint}
\end{equation}

\begin{widetext}
\section{From the correlator to the distribution}
\label{sec:CoToDist}
In this appendix we give the technical steps involved in going from 
the two-field correlator to the distribution $\cpd$, as discussed 
in Sec.~\ref{sec:ImplTFC}.  Beginning with Eq.~(\ref{eq:ProToCol}), 
setting 
${\bf k}_{2{\rm tot}}$ to be 
${\bf k}_{1{\rm tot}}$, 
and reorganizing, we obtain
\begin{eqnarray}
&&
Q^{2}\int d{\Delta r}_{11}\,d{\Delta r}_{22}\,d{\Delta r}_{12}\,
e^{-\frac{1}{2}{\Delta r}_{11d_{1}d_{2}}
    \sum_{\alpha=0}^{n}k_{1d_{1}}^{\alpha}k_{1d_{2}}^{\alpha}}\,
e^{-\frac{1}{2}{\Delta r}_{22d_{1}d_{2}}
    \sum_{\alpha=0}^{n}k_{2d_{1}}^{\alpha}k_{2d_{2}}^{\alpha}}\,
e^{            {\Delta r}_{12d_{1}d_{2}}
    \sum_{\alpha=0}^{n}k_{1d_{1}}^{\alpha}k_{2d_{2}}^{\alpha}}
\nonumber\\ &&\qquad\times
\int d{\bf x}\,  
e^{i{\bf k}_{{\rm tot}}\cdot{\bf x}}\,
\cpd({\bf x},{\Delta r}_{11},{\Delta r}_{22},{\Delta r}_{12})\,
=
\widetilde{\cw}(k_{1\rtr})\,
\widetilde{\cw}(k_{2\rtr})\,
\int\frac{d{\bf x}}{V}\,
e^{ i{\bf k}_{1{\rm tot}}\cdot{\bf x}}\,
e^{\frac{T}{\stiff_{0}}\,\,{\cal G}_{d_{1}d_{2}}({\bf x})\,
k_{1\rtr d_{1}}\cdot k_{2\rtr d_{2}}}. 
\label{eq:SimplifyCol}
\end{eqnarray}
Inserting the parametrization~(\ref{eq:parametrize}) for 
$\cpd$, performing the resulting integrations over 
${\Delta r}_{11}$, ${\Delta r}_{22}$ and ${\Delta r}_{12}$ 
on the LHS, exchanging factors of $\widetilde{\cw}$ for $\wt{\dsl}$  
via Eq.~(\ref{eq:shiftindistribution}), cancelling factors of $Q$, 
and identifying the Fourier transform 
\begin{equation}
\cpd({\bf q},\xi_{1}^{2},\xi_{2}^{2},{\bf y})
\equiv
\int d{\bf x}\,e^{i{\bf q}\cdot{\bf x}}\, 
\cpd({\bf x},\xi_{1}^{2},\xi_{2}^{2},{\bf y}),  
\end{equation}
we arrive at the folloinwg equation: 
\begin{eqnarray}
&&
\int 
d\xi_{1}^{2}\,\wt{\dsl}\left(\xi_{1}^{2}\right)\,
d\xi_{2}^{2}\,\wt{\dsl}\left(\xi_{2}^{2}\right)\,
d{\bf y}\,
\cpd({\bf k}_{1{\rm tot}},\xi_{1}^{2},\xi_{2}^{2},{\bf y})\,
e^{-\frac{1}{2}\xi_{1}^{2}k_{1}^{2}
   -\frac{1}{2}\xi_{2}^{2}k_{2}^{2}
   +\frac{T}{\stiff_{0}}\,\,{\cal G}_{d_{1}d_{2}}({\bf y})\,
                      k_{1d_{1}}\cdot k_{2d_{2}}}
\nonumber\\ 
&&\qquad=
\int 
d\xi_{1}^{2}\,\wt{\dsl}\left(\xi_{1}^{2}\right)\,
d\xi_{2}^{2}\,\wt{\dsl}\left(\xi_{2}^{2}\right)\,
e^{-\frac{1}{2}\xi_{1}^{2}k_{1\rtr}^{2}
   -\frac{1}{2}\xi_{2}^{2}k_{2\rtr}^{2}}
\int d{\bf y}\,
e^{ i{\bf k}_{1{\rm tot}}\cdot{\bf x}}\,
e^{\frac{T}{\stiff_{0}}\,\,
{\cal G}_{d_{1}d_{2}}({\bf y})\,
k_{1\rtr d_{1}}\cdot k_{2\rtr d_{2}}}. 
\end{eqnarray}
In order to clarify the content of this equation, we re-write the 
wave-vector dependence on the LHS in terms of replica-longitudinal 
(more precisely, $k_{\rtr}$) and replica-transverse 
(in fact ${\bf k}_{\rm tot}$) components 
[see Eqs.~(\ref{eq:tldecomp},\ref{eq:TotalMom})], 
noting that ${\bf k}_{1{\rm tot}}={\bf k}_{2{\rm tot}}$ 
and writing ${\bf q}$ for each, to obtain 
\begin{eqnarray}
&&
\int 
d\xi_{1}^{2}\,\wt{\dsl}\left(\xi_{1}^{2}\right)\,
d\xi_{2}^{2}\,\wt{\dsl}\left(\xi_{2}^{2}\right)\,
\int d{\bf y}\,
\cpd({\bf q},\xi_{1}^{2},\xi_{2}^{2},{\bf y})\,
e^{-\frac{1}{2}\xi_{1}^{2}k_{1\rtr}^{2}
   -\frac{1}{2}\xi_{2}^{2}k_{2\rtr}^{2}
   +\frac{T}{\stiff_{0}}\,\,{\cal G}_{d_{1}d_{2}}({\bf y})\,
                      k_{1\rtr d_{1}}\cdot k_{2\rtr d_{2}}}\,
e^{-\frac{{\bf q}\cdot{\bf q}}{2(1+n)}(\xi_{1}^{2}+\xi_{2}^{2})
   +\frac{T}{\stiff_{0}}\,\,
   {\cal G}_{d_{1}d_{2}}({\bf y})\,q_{d_{1}}q_{d_{2}}}
\nonumber\\ 
&&\qquad\quad=
\int 
d\xi_{1}^{2}\,\wt{\dsl}\left(\xi_{1}^{2}\right)\,
d\xi_{2}^{2}\,\wt{\dsl}\left(\xi_{2}^{2}\right)\,
e^{-\frac{1}{2}\xi_{1}^{2}k_{1\rtr}^{2}
   -\frac{1}{2}\xi_{2}^{2}k_{2\rtr}^{2}}
\int d{\bf y}\,
e^{ i{\bf q}\cdot{\bf y}}\,
e^{\frac{T}{\stiff_{0}}\,\,
  {\cal G}_{d_{1}d_{2}}({\bf y})\,
   k_{1\rtr d_{1}}\cdot k_{2\rtr d_{2}}}. 
\label{eq:SimpNewCol}
\end{eqnarray}
Next, we identify Laplace transformations with respect to 
$\xi_{1}^{2}$ and $\xi_{2}^{2}$, equate the entities being 
transformed on the LHS and RHS, and take the replica limit, 
thus arriving at 
\begin{equation}
e^{-\frac{1}{2}
\left(\xi_{1}^{2}+\xi_{2}^{2}\right)q^{2}}
\int d{\bf y}\,
\cpd({\bf q},\xi_{1}^{2},\xi_{2}^{2},{\bf y})\,
e^{\frac{T}{\stiff_{0}}\,\,{\cal G}_{d_{1}d_{2}}({\bf y})\,
                      k_{1\rtr d_{1}}\cdot k_{2\rtr d_{2}}}\,
e^{\frac{T}{\stiff_{0}}\,\,
   {\cal G}_{d_{1}d_{2}}({\bf y})\,q_{d_{1}}q_{d_{2}}}
=\int d{\bf y}\,
e^{i{\bf q}\cdot{\bf y}}\,
e^{\frac{T}{\stiff_{0}}\,\,
  {\cal G}_{d_{1}d_{2}}({\bf y})\,
   k_{1\rtr d_{1}}\cdot k_{2\rtr d_{2}}}. 
\label{eq:SimpNextCol}
\end{equation}
The next steps are to move the gaussian prefactor to the RHS, 
and to introduce the dummy variable $\bfgamma$, which takes on 
the values held by the second-rank tensor ${\cal G}({\bf y})$: 
\begin{eqnarray}
&&
\int d\bfgamma\,
e^{\frac{T}{\stiff_{0}}\,\,\bfgamma_{d_{1}d_{2}}\,
                      k_{1\rtr d_{1}}\cdot k_{2\rtr d_{2}}}
\int d{\bf y}\,
\delta\big(\bfgamma-{\cal G}({\bf y})\big)\,
\cpd({\bf q},\xi_{1}^{2},\xi_{2}^{2},{\bf y})\,
e^{\frac{T}{\stiff_{0}}\,\,
   {\cal G}_{d_{1}d_{2}}({\bf y})\,q_{d_{1}}q_{d_{2}}}
\nonumber\\ 
&&\qquad=
\int d\bfgamma\,
e^{\frac{T}{\stiff_{0}}\,\,
   \bfgamma_{d_{1}d_{2}}\,
    k_{1\rtr d_{1}}\cdot k_{2\rtr d_{2}}}
e^{\frac{1}{2}
\left(\xi_{1}^{2}+\xi_{2}^{2}\right)q^{2}}
\int d{\bf y}\,
e^{i{\bf q}\cdot{\bf y}}\,
\delta\big(\bfgamma-{\cal G}({\bf y})\big). 
\label{eq:SimpDum}
\end{eqnarray}
Again equating entites being transformed, this time being 
transformed with respect to $\bfgamma$, we find
\begin{eqnarray}
\int d{\bf y}\,
\delta\big(\bfgamma-{\cal G}({\bf y})\big)\,
\cpd({\bf q},\xi_{1}^{2},\xi_{2}^{2},{\bf y})\,
=e^{\frac{1}{2}\left(\xi_{1}^{2}+\xi_{2}^{2}\right)q^{2}}
\int d{\bf y}\,
e^{i{\bf q}\cdot{\bf y}}\,
\delta\big(\bfgamma-{\cal G}({\bf y})\big)\,
e^{-\frac{T}{\stiff_{0}}\,\,
   {\cal G}_{d_{1}d_{2}}({\bf y})\,q_{d_{1}}q_{d_{2}}}. 
\label{eq:NewDum}
\end{eqnarray}
Next, we equate entites being transformed as 
$\int d{\bf y}\,\delta\left(\bfgamma-{\cal G}({\bf y})\right)\cdots$, 
thus arriving at the Fourier transform of the reduced distribution: 
\begin{equation}
\cpd({\bf q},\xi_{1}^{2},\xi_{2}^{2},{\bf y})=
e^{\frac{1}{2}\left(\xi_{1}^{2}+\xi_{2}^{2}\right)q^{2}}\,
e^{i{\bf q}\cdot{\bf y}}\,
e^{-\frac{T}{\stiff_{0}}\,\,
   {\cal G}_{d_{1}d_{2}}({\bf y})\,q_{d_{1}}q_{d_{2}}}. 
\label{eq:NewTrans}
\end{equation}
Finally, we invert the Fourier transform to arrive at the 
reduced distribution: 
\begin{equation}
\cpd({\bf x},\xi_{1}^{2},\xi_{2}^{2},{\bf y})=
\int\dbar{\bf q}\,e^{-i{\bf q}\cdot{\bf x}}\,
\cpd({\bf q},\xi_{1}^{2},\xi_{2}^{2},{\bf y})=
\int\dbar{\bf q}\,e^{-i{\bf q}\cdot({\bf x}-{\bf y})}\,
e^{\frac{1}{2}
\left(\xi_{1}^{2}+\xi_{2}^{2}\right)
{\bf q}\cdot{\bf q}}\,
e^{-\frac{T}{\stiff_{0}}\,\,
   {\cal G}_{d_{1}d_{2}}({\bf y})\,q_{d_{1}}q_{d_{2}}}. 
\label{eq:UnFourier}
\end{equation}
Convergence of this inversion is furnished by the cut-off nature 
of the ${\bf q}$ integration.  Inserting this formula into the 
parametrization~(\ref{eq:parametrize}) we arrive at a formula 
for the full distribution: 
\begin{eqnarray}
&&
\cpd({\bf x},{\Delta r}_{11},{\Delta r}_{22},{\Delta r}_{12})
=\int
d\xi_{1}^{2}\,\wt{\dsl}\left(\xi_{1}^{2}\right)\,
d\xi_{2}^{2}\,\wt{\dsl}\left(\xi_{2}^{2}\right)\,
\delta\big({\Delta r}_{11}-\xi_{1}^{2}\openone\big)\,
\delta\big({\Delta r}_{22}-\xi_{2}^{2}\openone\big)
\nonumber\\
&&
\qquad\qquad\qquad
\times
\int\frac{{d{\bf y}}}{V}\,
\left\{
\int\dbar{\bf q}\,e^{-i{\bf q}\cdot({\bf x}-{\bf y})}\,
e^{\frac{1}{2}\left(\xi_{1}^{2}+\xi_{2}^{2}\right)q^{2}}
e^{-\frac{T}{\stiff_{0}}\,\,
   {\cal G}_{d_{1}d_{2}}({\bf y})\,q_{d_{1}}q_{d_{2}}}
\right\}
\delta\big({\Delta r}_{12}-({T}/{\stiff_{0}})\,
        {\cal G}({\bf y})\big). 
\label{eq:useparam}
\end{eqnarray}
\end{widetext}


\end{document}